 \documentclass[pmlr,10pt]{jmlr} 

\usepackage{booktabs}
\usepackage{multirow}
\usepackage{colortbl}
\usepackage{cleveref}
\usepackage{wrapfig}
\usepackage{pifont}

\usepackage{siunitx}

\usepackage[switch]{lineno}

\theorembodyfont{\upshape}
\theoremheaderfont{\scshape}
\theorempostheader{:}
\theoremsep{\newline}

 \title[On the challenges of detecting MCI using EEG in the wild]{On the challenges of detecting MCI using EEG in the wild}

\author{%
\Name{Aayush Mishra} \Email{amishr24@jhu.edu}\\
\addr Johns Hopkins University
\AND
\Name{David Joffe} \Email{david@joffemail.com}\\
\addr Wavi Co.
\AND
\Name{Sankara Surendra Telidevara} \Email{stelide1@jhu.edu}\\
\addr Johns Hopking University
\AND
\Name{David S Oakley} \Email{davido@wavimed.com}\\
\addr Wavi Co.
\AND
\Name{Anqi Liu}
\Email{aliu@cs.jhu.edu}\\
\addr Johns Hopkins University
}

\usepackage{color}

\newcommand\filledcircb{\ensuremath{{\color[HTML]{2600FF}\bullet}}}
\newcommand\filledcircr{\ensuremath{{\color[HTML]{FF0000}\bullet}}}

\begin{document}

\maketitle

\begin{abstract}
Recent studies have shown promising results in the detection of Mild Cognitive Impairment (MCI) using easily accessible Electroencephalogram (EEG) data which would help administer early and effective treatment for dementia patients. However, the reliability and practicality of such systems remains unclear. 
In this work, we investigate the potential limitations and challenges in developing a robust MCI detection method using 
two contrasting datasets: 1) CAUEEG, collected and annotated by expert neurologists in controlled settings and 2) GENEEG, a new dataset collected and annotated in general practice clinics, a setting where routine MCI diagnoses are typically made. We find that training on small datasets, as is done by most previous works, tends to produce high variance models that make overconfident predictions, and are unreliable in practice. Additionally, distribution shifts between datasets make cross-domain generalization challenging. 
Finally, we show that MCI detection using EEG may suffer from fundamental limitations because of the overlapping nature of feature distributions with control groups. 
We call for more effort in high-quality data collection in actionable settings (like general practice clinics) to make progress towards this salient goal of non-invasive MCI detection.

\end{abstract}


\vspace{5pt}
\paragraph*{Data and Code Availability}
Details about CAUEEG dataset can be found in \citep{kim2023deep}. Our new GENEEG dataset can be accessed at \href{https://huggingface.co/datasets/aamixsh/GENEEG}{huggingface}. Code used in this study can be found at \href{https://github.com/aamixsh/eeg_mci}{github}.

\paragraph*{Institutional Review Board (IRB)}
Our clinical data acquisition and data analysis was conducted under IRB approval (IRB00348520) in the Johns Hopkins University. 

\section{Introduction}
\label{sec:intro}

Every year, more than 10 million people develop Dementia, according to a report from the World Health Organization \citep{WHO}. More than half of them are from middle to low-income countries. It is one of the leading causes of disability and dependency among old people. Around 60-80\% cases of dementia are caused by Alzheimer’s Disease \citep{ALZ}. 
Mild Cognitive Impairment (MCI) is a condition in which people start having problems with thinking and memory. 
Studies have shown that people with MCI are at a higher risk of developing Alzheimer's or a related form of Dementia \citep{campbell2013risk}. Without a cure, early detection of MCI is one of the most effective ways to delay the progression of this condition. 

Most initial MCI diagnoses are made in the general practice setting, where the earliest indicators are seen before specialist referrals. Diagnoses in these settings are typically made 
through patient/family interviews and mental status tests like Montreal Cognitive Assessment (MoCA) or Mini-Mental Status Examination (MMSE). Many MCI phenotypes can be delayed, prevented, or reversed with interventions that target sleep, cardio health, mood, and hearing loss, to name a few. The general practitioner is often better equipped to offer these solutions than neurological specialists, particularly in the early stages. General practice clinics could benefit from physiological information for early detection and hence intervention.

The medical diagnosis of MCI is difficult because the condition usually occurs in older people ($> 60$ years) and the symptoms can often be regarded as artifacts of aging. Many expensive, inaccessible, and sometimes invasive tests are required to detect MCI which poses a challenge to the hopes of early detection. On the other hand, electroencephalogram (EEG) recording systems are simpler, non-invasive, and cheap. EEG signals have been shown to be a possible disease marker \citep{maitin2022survey, hosseini2020review, rasheed2020machine, lee2022real}. Recent work \citep{yang2019m, rossini2022early} have also shown that machine learning methods can be used to model EEG signals as markers in detecting MCI. Although these studies offer a promising future of easier disease management, they are typically performed on small neurologist-labeled datasets that make the application and generalizability of their findings unreliable. In this work, we investigate this problem with a variety of datasets and models. In summary:

\begin{figure*}[t]
    \centering
    \includegraphics[width=0.99\linewidth]{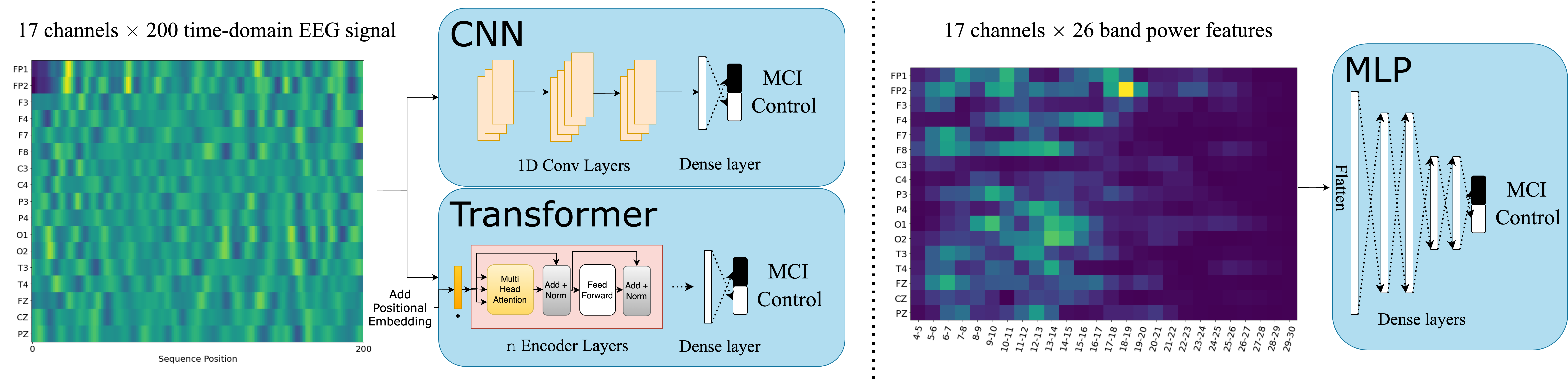}
	\vspace{-5pt}
	\caption{\small{An overview of the neural networks used in this study. The CNN and Transformer models are used to model time-domain EEG signals while an MLP is used to model frequency domain transformed data.}}\vspace{-10pt}
    \label{fig:nns}
\end{figure*}
\begin{itemize}
    \item We present a \textbf{new dataset (GENEEG)}, collected and annotated in general practice clinics. The objective is to target early MCI detection in a setting that replicates how MCI diagnoses are made in the real world, rather than the laboratory setting previously studied.
    \item We evaluate Deep Learning models (MLP, CNN and Transformer) with \textbf{reliability metrics} and \textbf{cross-domain testing} to illustrate the untrustworthiness of popular models in this problem. 
    \item We find that there exists substantial feature overlap between MCI and Control patients, reflecting the potential inherent \textbf{aleatoric uncertainty} that is hard to reduce in this problem.
\end{itemize}

\section{Related Work}
\label{sec:related}

EEG signals have long been proposed as possible markers for Alzheimer's Disease (AD)~\citep{jeong2004eeg, dauwels2010diagnosis, melissant2005method, safi2021early}. It is hard to distinguish MCI symptoms from AD, but many recent works also study the detection of MCI using EEG signals. These range from several studies that focus on feature extraction and then modeling \citep{aljalal2024mild, siuly2020new, pirrone2022eeg} to other studies that use end-to-end training of neural networks \citep{morabito2016deep, alvi2022long}. Other recent work like \citep{ieracitano2020novel} utilizes multi-modal models to leverage features from different representations of the same input. 

However, many of these studies focus on their specific data and patient cohorts and involve design choices that may not be generalized to other tasks. For example, \cite{cseker2024deep} use models like 2D Convolutional Neural Nets (CNNs) for EEG signals that are sequences with primarily temporal dependencies (1D). 
\cite{kashefpoor2016automatic} use feature correlation as a metric to select features, while spurious correlation is a well-known problem in machine learning which question the efficacy of such a decision. 
In this work, we explore and evaluate three different model architectures whose design is influenced by the nature of data. \citet{kim2023deep} recently published a similar relevant study which attempts to evaluate multiple models on their new CAUEEG dataset. In our work, we use this dataset along with our new GENEEG dataset to show the impact of distribution shifts on performance.

\section{Data}
\label{sec:data}

\subsection{Datasets}
\label{subsec:data}

\subsubsection*{GENEEG (\textbf{new})}
\label{subsec:geneeg}
We introduce a new EEG dataset (GENEEG) of 28 MCI and 35 Control patients collected at general practice clinics and labeled by general practitioners. 
The purpose of GENEEG is to provide a test bed for the applicability of MCI detection in the real world setting of typical diagnosis. This is in contrast to other datasets for MCI detection, which are usually well-vetted and labeled by expert neurologists.

\begin{wrapfigure}{r}{0.35\textwidth}
\vspace{-20pt}
    \centering
    \includegraphics[width=0.90\linewidth]{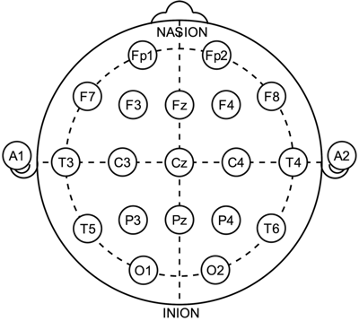}
	\vspace{-5pt}
	\caption{\small{Topographical map of the active EEG channels. T5 and T6 were not used in the study.}}\vspace{-20pt}
    \label{fig:head}
\end{wrapfigure}

\paragraph{Participant information} Data from a total of 28 patients labelled MCI and 35 patients labelled Normal (Control) was collected. The diagnoses were made by clinicians in general clinics using standard procedure \citep{petersen2018practice}. This represents a heterogeneous sample of real-life MCI diagnoses. Participants from both groups had the same mean age of 73.4 years, with standard deviations of 4.7 and 9.9 for the Control and MCI groups respectively. It is noteworthy that this is a small dataset which can introduce reliability concerns about results and inferences, as we see in \autoref{sec:exp}.

\paragraph{Data acquisition} 
The control and MCI EEG data files were acquired using a standard 19 channel EEG acquisition system during the administration of a 4 minute eyes-closed P300 protocol at a sampling rate of 250 Hz, using the standard (international 10-20 system) electrode montage. 
\autoref{fig:head} illustrates the positioning of the electrodes. Here, A1 and A2 are linked-ear references. We excluded the T5 and T6 channels for technical reasons. 
As the CAUEEG dataset had a different sampling rate of 200 Hz and used common average referencing, we subsampled GENEEG to create the same sampling frequency for the purpose of standardization.

\subsubsection*{CAUEEG}
\label{subsec:caueeg}
\citet{kim2023deep} introduced CAUEEG, which contains EEG data of 1155 patients collected at Chung-Ang University in South Korea. 
This data was labelled using standard criteria and psychological tests to give at least one of 28 different diagnoses. 
For this work, we focus on EEG recordings with labels of either MCI, Dementia or Normal (Control). After standard artefacting (described in \autoref{subsec:datadesign}), there were 385 MCI patients, 292 Dementia patients and 441 Control patients remaining. This is a high quality big dataset suitable for deep learning, which we used to study MCI detection in clinically controlled settings. 

\paragraph{Why use both datasets?} While GENEEG offers a more realistic data collection setting, it is relatively small. Models trained on small datasets tend to suffer from problems like over-fitting and high performance variance (as we see in our experiments). On the other hand, CAUEEG offers a large-scale expert-vetted dataset with multiple conditions (MCI and Dementia) that can be used to measure the impact of distribution shifts and reliability across domains. We use both datasets to explore the nuances that remain unexplored in previous studies.

\subsection{Pre-Processing}
\label{subsec:datadesign}

\paragraph{Filtering and Artifact Removal} Both datasets were first band passed between 5 to 20 Hz to remove low and high-frequency artifacts followed by manual artifacting. This step marks the segments of the EEG signal that will be excluded from analysis in order to increase the data yield.

\begin{figure}[t]
    \centering
    \includegraphics[height=1.5in,trim=1cm 0cm 1cm 1cm,clip=true]{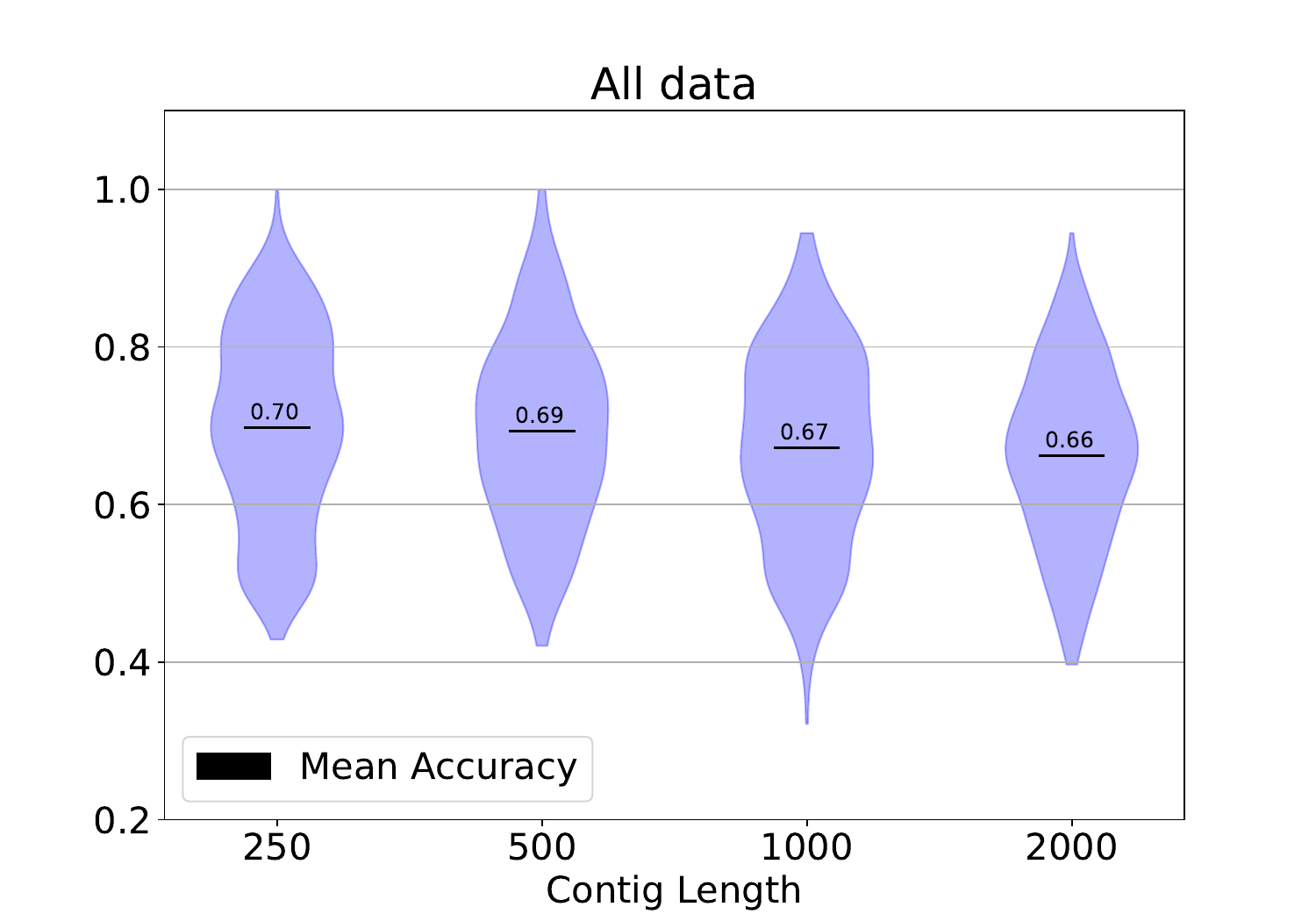}
    \includegraphics[height=1.5in,trim=0.9cm 0cm 1cm 1cm,clip=true]{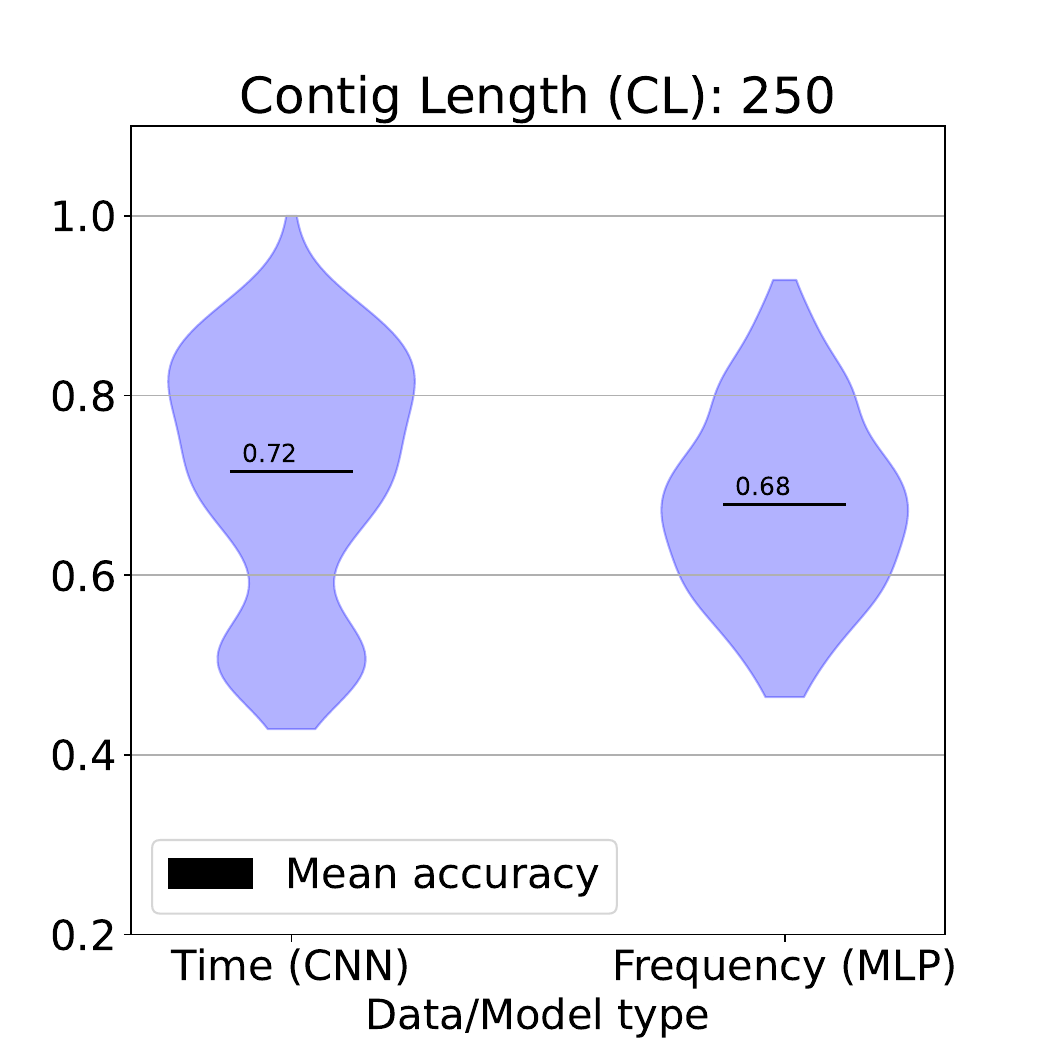}
    \includegraphics[height=1.5in,trim=1cm 0cm 1cm 1cm,clip=true]{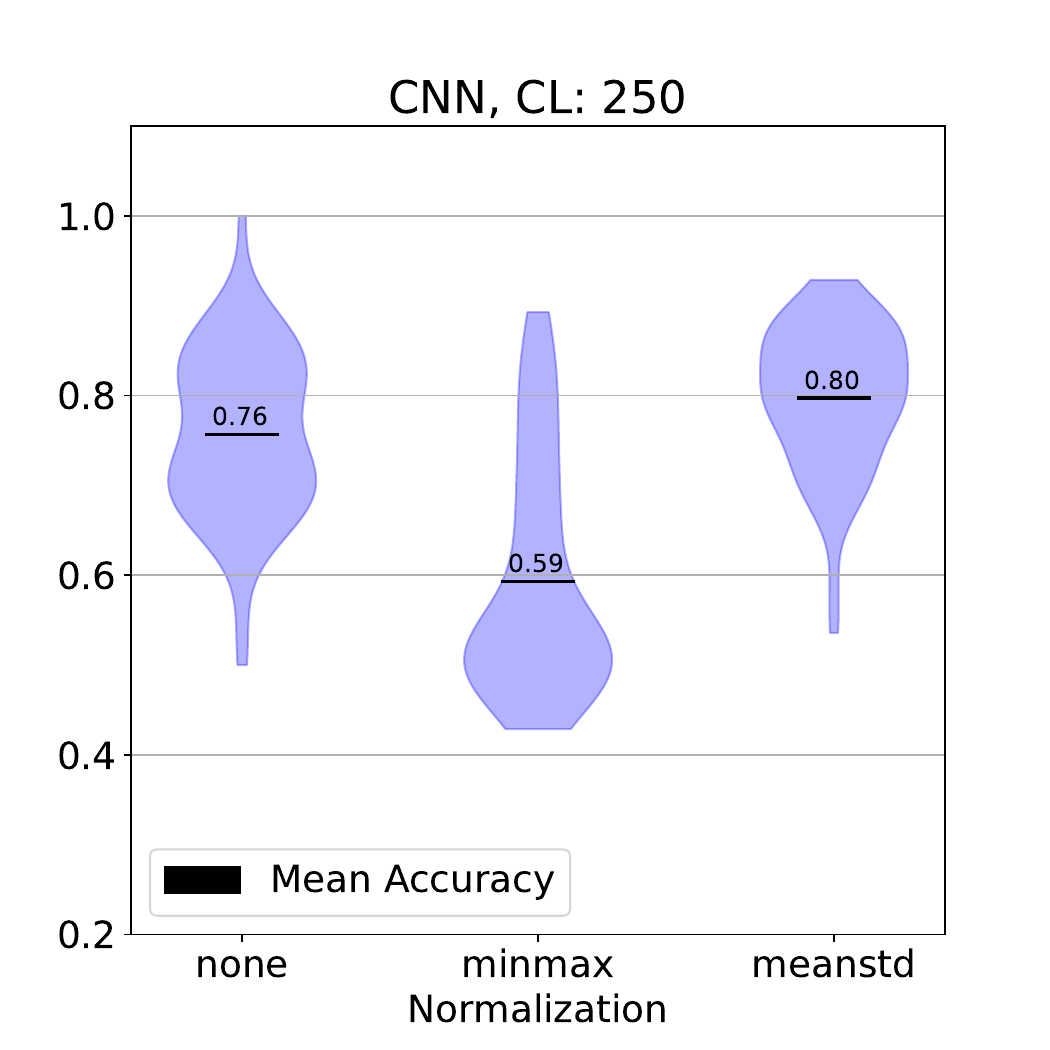}
    \includegraphics[height=1.5in,trim=0.9cm 0cm 1cm 1cm,clip=true]{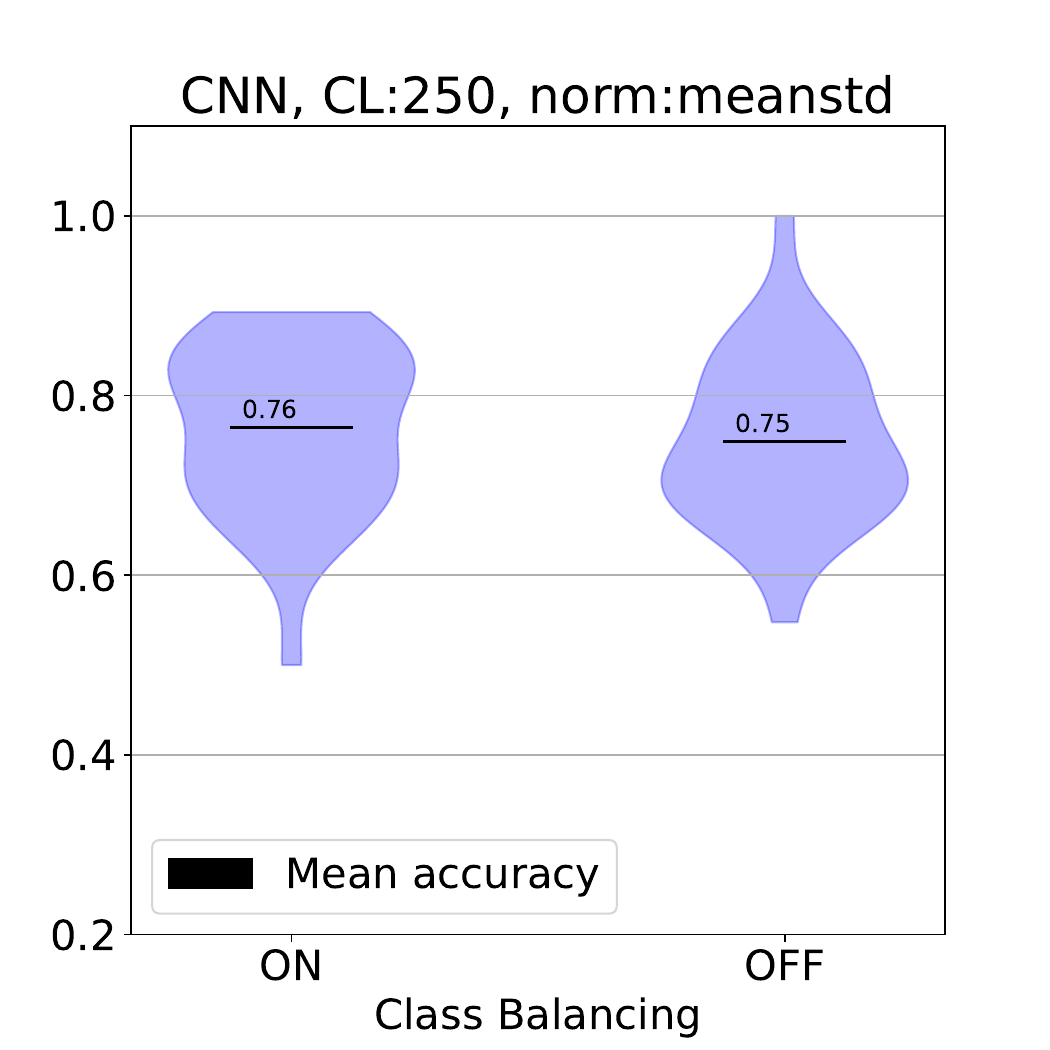}\\
    (a) \hspace{125pt} (b) \hspace{85pt} (c) \hspace{85pt} (d)
	\vspace{-5pt}
	\caption{\small{In our ablation experiments, we found that models tend to perform better with smaller contig lengths, on time-domain data with standardization. The effect of class balancing was negligible. This is how we settled on data configuration (contig length, normalization and class balancing) before our final experiments.}}\vspace{-10pt}
    \label{fig:ablation}
\end{figure}

\paragraph{Splitting into fixed-length \textit{Contigs}} To exclude segments from long EEG signals that contain noisy artifacts, we split them into 1 second (200 time steps) fixed length segments called \textit{contigs} after experimenting with different contig lengths. We experiment with multiple contig lengths (250, 500, 1000, 2000) in GENEEG (due to small dataset size) on CNN and MLP models, and found that the smallest 250 length contigs performed best overall (\autoref{fig:ablation}).
We then use subsampling from this signal to convert it into 200 length sequences we finally use to compare with CAUEEG. With that, we settle on the following dataset statistics for our performance and reliability evaluation experiments:
\begin{enumerate}
    \item CAUEEG: 385 MCI patients, mean contigs per patient $\approx$ 332 (std $\approx$ 149); 441 Control patients, mean contigs per patient $\approx$ 341 (std $\approx$ 130); 292 Dementia patients, mean contigs per patient $\approx$ 358 (std $\approx$ 161). Total contigs: 382832.
    \item GENEEG: 28 MCI patients, mean contigs per patient $\approx$ 186 (std $\approx$ 31); 35 Control patients, mean contigs per patient $\approx$ 223 (std $\approx$ 11). Total contigs: 13009.
\end{enumerate}

\begin{figure*}[t]
    \centering
    \includegraphics[width=0.99\linewidth]{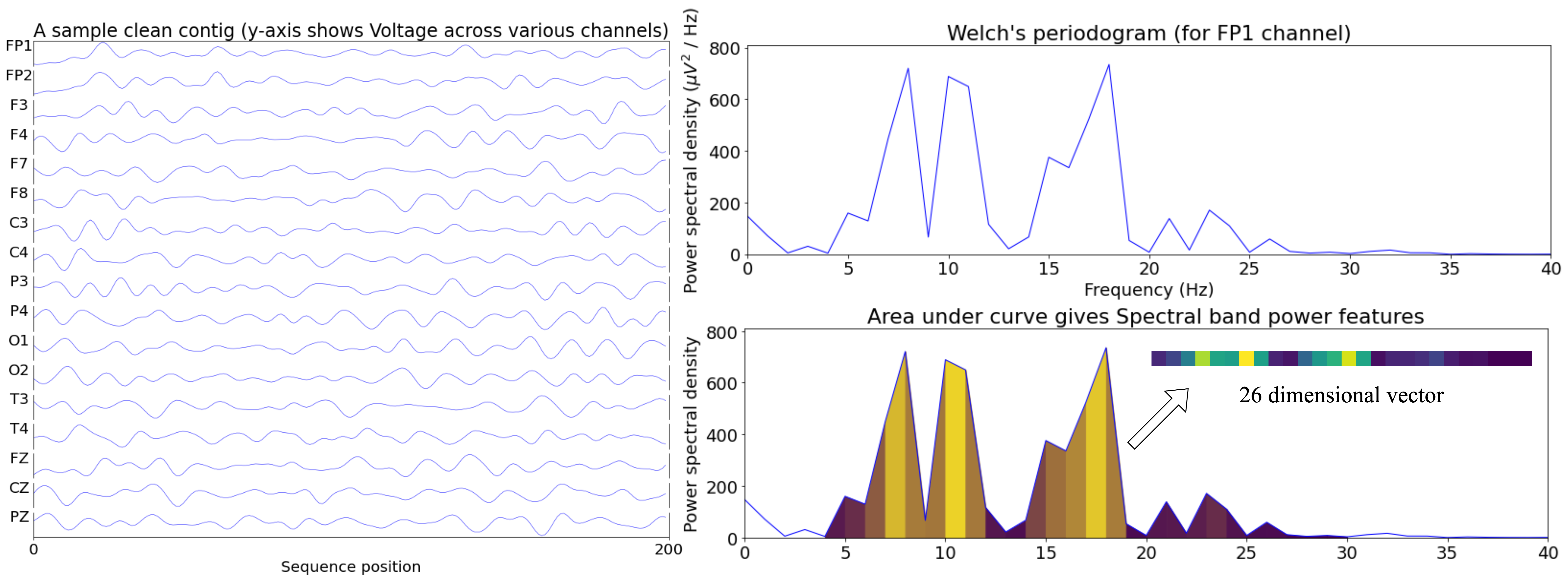}
    \caption{\small{Conversion of time-domain EEG signals to frequency domain spectral features by discretizing the power density spectrum. For bands of size 1 Hz from 4-30 Hz, we get 26 band power features per channel.}}\vspace{-10pt}
    \label{fig:rts}
\end{figure*}

\paragraph{Time Domain vs Frequency Domain}
Many prior work transform the time-domain EEG signals to the frequency domain using the Fourier transform~\cite{}. Spectral density gives a measure of how much each frequency component contributes to the signal. 
While it is easier to model the spectral densities due to the lower dimensionality, it results in a loss of phase information present in the time-domain data. We study data from both time and frequency domains. 
Rather than using the standard $\delta, \theta, \alpha, \beta$ and $\gamma$ bands, we choose a finer resolution 
1 Hz bands
ranging from 4 to 30 Hz (26 features), as shown in 
\autoref{fig:rts}.

\paragraph{Class balancing and Normalization}
Class imbalance often causes models to be biased towards the majority class. We study the effect of randomly subsampling data from the majority class to maintain training set balance. We also test standardization and normalization as it is beneficial when training neural networks \citep{huang2023normalization}. In general, we find that data balancing has no significant impact (so we chose to leave it unbalanced), and standardization of data usually yields better performance on our datasets (\autoref{fig:ablation}) for our final experiments. \autoref{tab:cmacc} in the appendix shows detailed results of this ablation study.

\section{Experimental Setup}
\label{sec:setup}

\subsection{Models}
\label{subsec:models}

Neural networks are universal approximators \citep{cybenko1989approximation}. We follow previous work \citep{kim2023deep, ieracitano2020novel} to
use various neural networks in this study. 

\paragraph{Architectures}
Motivated by the different properties of the time and frequency domains, we choose different neural networks for each with appropriate inductive biases.
\begin{enumerate}
    \item We have temporal dependencies and patterns in the time domain data. Therefore, we apply \textbf{CNN} and \textbf{Transformer (TF)} backends
    to learn representations before final classification using dense layers as shown in \autoref{fig:nns}. 
    CNNs are widely used to capture local patterns which may be translation invariant. Transformers can model even long-range dependencies.
    \item Each location in the $17 \times 26$ frequency domain input represents one unique feature, whose location is immutable. 
    Therefore, we use a standard Multi-Layer Perceptron (\textbf{MLP} or feed-forward neural network) as shown in \autoref{fig:nns}.
\end{enumerate} 

\paragraph{Training Configuration and Hyper-parameters}
Neural Networks are known to be sensitive to training configurations. We performed a grid search to find good training configurations and hyperparameters like
number of convolution filters, number of attention heads, 
learning rates, etc. 
More details can be found in Appendix \autoref{app:grid}. 

\subsection{Post Training}
\label{subsec:post}

\paragraph{Prediction}
\label{subsec:pred}

Our models are trained to classify contigs instead of patients. These contig-level predictions need to be aggregated to a patient-level prediction. We employ a simple \textbf{majority voting strategy} (counting the number of contigs predicted in each class) to give patient-level predictions. 
We can also use a validation set to pick prediction probability threshold that performs best, however it is likely that the threshold is biased by the limited validation set. We defer the exploration of other aggregation strategies to future work.

\paragraph{Evaluation}\label{sec:eval} For performance, we evaluate our models on Recall and Precision on the class of interest (MCI/Dementia).
Instead of reporting mean values across samples (micro), we report individual class averaged (macro) values to take class imbalance into account which sets the baselines at \textbf{0.5}.
We also measure \textit{Expected Calibration Error} (ECE) \citep{guo2017calibration} and \textit{Uncertainty} $=\frac{p * (1 - p)}{0.5}$ (for correct and incorrect predictions separately) to evaluate reliability. A reliable model knows when it does not know. Therefore, it should have low calibration errors, high uncertainty for incorrect predictions and low uncertainty for correct predictions. We use a 75:25 train/test split at patient level to evaluate these models. The process is repeated 30 times to make sure that the evaluation is reliable.

\section{Experimental Results}
\label{sec:exp}

\begin{table}[]
\caption{\small{Condition specific (MCI/Dementia), as well as macro (average of condition and control group) evaluation metrics (recall and precision) for different dataset, task and model combinations.
}}
\vspace{3pt}
\label{tab:main_table}
\resizebox{\textwidth}{!}{
\begin{tabular}{|c|cc|
>{\columncolor[HTML]{FFFFFF}}c 
>{\columncolor[HTML]{FFFFFF}}c 
>{\columncolor[HTML]{FFFFFF}}c 
>{\columncolor[HTML]{FFFFFF}}c |
>{\columncolor[HTML]{FFFFFF}}c 
>{\columncolor[HTML]{FFFFFF}}c |}
\hline
                                                                               & \multicolumn{2}{c|}{Dataset $\rightarrow$}                                                                 & \multicolumn{4}{c|}{\cellcolor[HTML]{9B9B9B}\textbf{CAUEEG}}                                                                                                                                                                              & \multicolumn{2}{c|}{\cellcolor[HTML]{9B9B9B}\textbf{GENEEG}}                                      \\ \cline{2-9} 
                                                                               & \multicolumn{2}{c|}{Task $\rightarrow$}                                                                    & \multicolumn{2}{c|}{\cellcolor[HTML]{EFEFEF}MCI v Control}                                                                            & \multicolumn{2}{c|}{\cellcolor[HTML]{EFEFEF}Dementia v Control}                                   & \multicolumn{2}{c|}{\cellcolor[HTML]{EFEFEF}MCI v Control}                                        \\ \cline{2-9} 
\multirow{-3}{*}{\begin{tabular}[c]{@{}c@{}}Model\\ $\downarrow$\end{tabular}} & \multicolumn{1}{c|}{}                                                    &                                 & \multicolumn{1}{c|}{\cellcolor[HTML]{EFEFEF}MCI}                  & \multicolumn{1}{c|}{\cellcolor[HTML]{EFEFEF}Macro}                & \multicolumn{1}{c|}{\cellcolor[HTML]{EFEFEF}Dementia}             & \cellcolor[HTML]{EFEFEF}Macro & \multicolumn{1}{c|}{\cellcolor[HTML]{EFEFEF}MCI}                  & \cellcolor[HTML]{EFEFEF}Macro \\ \hline
\cellcolor[HTML]{C0C0C0}                                                       & \multicolumn{1}{c|}{\cellcolor[HTML]{C0C0C0}}                            & \cellcolor[HTML]{EFEFEF}Contig  & \multicolumn{1}{c|}{\cellcolor[HTML]{FFFFFF}0.52 (0.03)}          & \multicolumn{1}{c|}{\cellcolor[HTML]{FFFFFF}0.58 (0.01)}          & \multicolumn{1}{c|}{\cellcolor[HTML]{FFFFFF}0.59 (0.04)}          & 0.70 (0.02)                   & \multicolumn{1}{c|}{\cellcolor[HTML]{FFFFFF}0.50 (0.10)}          & 0.63 (0.04)                   \\ \cline{3-9} 
\cellcolor[HTML]{C0C0C0}                                                       & \multicolumn{1}{c|}{\multirow{-2}{*}{\cellcolor[HTML]{C0C0C0}Recall}}  & \cellcolor[HTML]{EFEFEF}Patient & \multicolumn{1}{c|}{\cellcolor[HTML]{FFFFFF}0.53 (0.07)}          & \multicolumn{1}{c|}{\cellcolor[HTML]{FFFFFF}0.67 (0.02)}          & \multicolumn{1}{c|}{\cellcolor[HTML]{FFFFFF}0.66 (0.07)}          & 0.79 (0.03)                   & \multicolumn{1}{c|}{\cellcolor[HTML]{FFFFFF}0.55 (0.20)}          & 0.71 (0.09)                   \\ \cline{2-9} 
\cellcolor[HTML]{C0C0C0}                                                       & \multicolumn{1}{c|}{\cellcolor[HTML]{C0C0C0}}                            & \cellcolor[HTML]{EFEFEF}Contig  & \multicolumn{1}{c|}{\cellcolor[HTML]{FFFFFF}0.56 (0.02)}          & \multicolumn{1}{c|}{\cellcolor[HTML]{FFFFFF}0.59 (0.01)}          & \multicolumn{1}{c|}{\cellcolor[HTML]{FFFFFF}0.67 (0.03)}          & 0.71 (0.02)                   & \multicolumn{1}{c|}{\cellcolor[HTML]{FFFFFF}0.58 (0.07)}          & 0.64 (0.04)                   \\ \cline{3-9} 
\multirow{-4}{*}{\cellcolor[HTML]{C0C0C0}\textbf{MLP}}                         & \multicolumn{1}{c|}{\multirow{-2}{*}{\cellcolor[HTML]{C0C0C0}Precision}} & \cellcolor[HTML]{EFEFEF}Patient & \multicolumn{1}{c|}{\cellcolor[HTML]{FFFFFF}0.71 (0.04)}          & \multicolumn{1}{c|}{\cellcolor[HTML]{FFFFFF}0.69 (0.02)}          & \multicolumn{1}{c|}{\cellcolor[HTML]{FFFFFF}0.86 (0.04)}          & 0.83 (0.03)                   & \multicolumn{1}{c|}{\cellcolor[HTML]{FFFFFF}0.79 (0.15)}          & 0.76 (0.09)                   \\ \hline
\cellcolor[HTML]{C0C0C0}                                                       & \multicolumn{1}{c|}{\cellcolor[HTML]{C0C0C0}}                            & \cellcolor[HTML]{EFEFEF}Contig  & \multicolumn{1}{c|}{\cellcolor[HTML]{FFFFFF}0.50 (0.08)}          & \multicolumn{1}{c|}{\cellcolor[HTML]{FFFFFF}0.58 (0.02)}          & \multicolumn{1}{c|}{\cellcolor[HTML]{FFFFFF}0.61 (0.07)}          & 0.64 (0.02)                   & \multicolumn{1}{c|}{\cellcolor[HTML]{FFFFFF}0.31 (0.11)}          & 0.64 (0.05)                   \\ \cline{3-9} 
\cellcolor[HTML]{C0C0C0}                                                       & \multicolumn{1}{c|}{\multirow{-2}{*}{\cellcolor[HTML]{C0C0C0}Recall}}  & \cellcolor[HTML]{EFEFEF}Patient & \multicolumn{1}{c|}{\cellcolor[HTML]{FFFFFF}0.55 (0.12)}          & \multicolumn{1}{c|}{\cellcolor[HTML]{FFFFFF}0.63 (0.03)}          & \multicolumn{1}{c|}{\cellcolor[HTML]{FFFFFF}0.68 (0.10)}          & 0.73 (0.03)                   & \multicolumn{1}{c|}{\cellcolor[HTML]{FFFFFF}0.26 (0.14)}          & 0.63 (0.07)                   \\ \cline{2-9} 
\cellcolor[HTML]{C0C0C0}                                                       & \multicolumn{1}{c|}{\cellcolor[HTML]{C0C0C0}}                            & \cellcolor[HTML]{EFEFEF}Contig  & \multicolumn{1}{c|}{\cellcolor[HTML]{FFFFFF}0.56 (0.03)}          & \multicolumn{1}{c|}{\cellcolor[HTML]{FFFFFF}0.58 (0.02)}          & \multicolumn{1}{c|}{\cellcolor[HTML]{FFFFFF}0.58 (0.04)}          & 0.65 (0.02)                   & \multicolumn{1}{c|}{\cellcolor[HTML]{FFFFFF}\textbf{0.86 (0.09)}} & \textbf{0.78 (0.05)}          \\ \cline{3-9} 
\multirow{-4}{*}{\cellcolor[HTML]{C0C0C0}\textbf{TF}}                          & \multicolumn{1}{c|}{\multirow{-2}{*}{\cellcolor[HTML]{C0C0C0}Precision}} & \cellcolor[HTML]{EFEFEF}Patient & \multicolumn{1}{c|}{\cellcolor[HTML]{FFFFFF}0.65 (0.08)}          & \multicolumn{1}{c|}{\cellcolor[HTML]{FFFFFF}0.65 (0.04)}          & \multicolumn{1}{c|}{\cellcolor[HTML]{FFFFFF}0.69 (0.09)}          & 0.74 (0.04)                   & \multicolumn{1}{c|}{\cellcolor[HTML]{FFFFFF}\textbf{0.87 (0.34)}} & 0.75 (0.19)                   \\ \hline
\cellcolor[HTML]{C0C0C0}                                                       & \multicolumn{1}{c|}{\cellcolor[HTML]{C0C0C0}}                            & \cellcolor[HTML]{EFEFEF}Contig  & \multicolumn{1}{c|}{\cellcolor[HTML]{FFFFFF}\textbf{0.55 (0.04)}} & \multicolumn{1}{c|}{\cellcolor[HTML]{FFFFFF}\textbf{0.62 (0.01)}} & \multicolumn{1}{c|}{\cellcolor[HTML]{FFFFFF}\textbf{0.64 (0.05)}} & \textbf{0.74 (0.02)}          & \multicolumn{1}{c|}{\cellcolor[HTML]{FFFFFF}\textbf{0.62 (0.12)}} & \textbf{0.70 (0.05)}          \\ \cline{3-9} 
\cellcolor[HTML]{C0C0C0}                                                       & \multicolumn{1}{c|}{\multirow{-2}{*}{\cellcolor[HTML]{C0C0C0}Recall}}  & \cellcolor[HTML]{EFEFEF}Patient & \multicolumn{1}{c|}{\cellcolor[HTML]{FFFFFF}\textbf{0.60 (0.06)}} & \multicolumn{1}{c|}{\cellcolor[HTML]{FFFFFF}\textbf{0.70 (0.03)}} & \multicolumn{1}{c|}{\cellcolor[HTML]{FFFFFF}\textbf{0.70 (0.06)}} & \textbf{0.81 (0.03)}          & \multicolumn{1}{c|}{\cellcolor[HTML]{FFFFFF}\textbf{0.67 (0.20)}} & \textbf{0.77 (0.10)}          \\ \cline{2-9} 
\cellcolor[HTML]{C0C0C0}                                                       & \multicolumn{1}{c|}{\cellcolor[HTML]{C0C0C0}}                            & \cellcolor[HTML]{EFEFEF}Contig  & \multicolumn{1}{c|}{\cellcolor[HTML]{FFFFFF}\textbf{0.60 (0.02)}} & \multicolumn{1}{c|}{\cellcolor[HTML]{FFFFFF}\textbf{0.62 (0.01)}} & \multicolumn{1}{c|}{\cellcolor[HTML]{FFFFFF}\textbf{0.73 (0.03)}} & \textbf{0.75 (0.02)}          & \multicolumn{1}{c|}{\cellcolor[HTML]{FFFFFF}0.66 (0.05)}          & 0.71 (0.05)                   \\ \cline{3-9} 
\multirow{-4}{*}{\cellcolor[HTML]{C0C0C0}\textbf{CNN}}                         & \multicolumn{1}{c|}{\multirow{-2}{*}{\cellcolor[HTML]{C0C0C0}Precision}} & \cellcolor[HTML]{EFEFEF}Patient & \multicolumn{1}{c|}{\cellcolor[HTML]{FFFFFF}\textbf{0.73 (0.04)}} & \multicolumn{1}{c|}{\cellcolor[HTML]{FFFFFF}\textbf{0.71 (0.03)}} & \multicolumn{1}{c|}{\cellcolor[HTML]{FFFFFF}\textbf{0.86 (0.04)}} & \textbf{0.84 (0.02)}          & \multicolumn{1}{c|}{\cellcolor[HTML]{FFFFFF}0.82 (0.13)}          & \textbf{0.80 (0.10)}          \\ \hline
\end{tabular}
}
\end{table}

\autoref{tab:main_table} reports Recall and Precision for all datasets and model combinations. 
Note that we show these results on the best performing training configuration chosen using grid search. We can make several observations from these numbers.

\paragraph{CNN outperforms other Models} We find that our simple 1D CNN model outperforms Transformer and MLP models across all datasets in terms of Recall (Note that as we measure macro recall, i.e. average of recall for individual classes, it is equal to macro accuracy). This points to the importance of local translation invariant temporal features. While transformers are equipped with long-range dependency modeling, these long-range patterns might be noisy and less useful than the local features that the CNN captures. On the other hand, the MLP model trained on spectral features misses out on \textit{phase information} present in time-domain signals, which makes it perform worse. It is noteworthy that the Transformer model gets a higher precision score in GENEEG but has low recall (=accuracy) for the MCI class. This means that the models mostly predict the larger Control class and only predicts MCI for very confident cases. But this makes it miss out on a large number of MCI cases, making it overall worse than the CNN model.

\begin{figure}[t]
    \centering
    \includegraphics[width=0.245\linewidth,trim=0.15cm 0cm 1cm 1cm,clip=true]{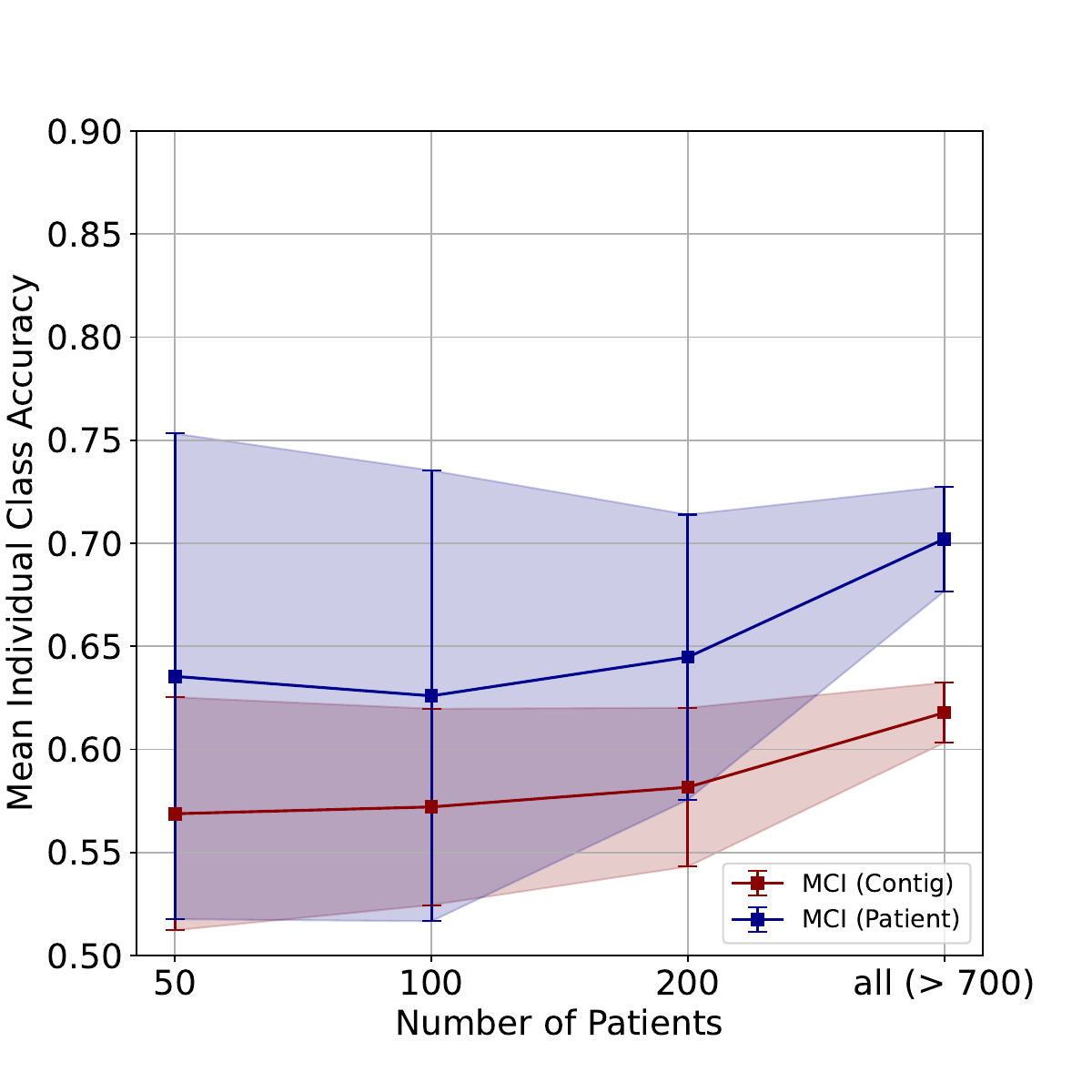}
    \includegraphics[width=0.245\linewidth,trim=0.15cm 0cm 1cm 1cm,clip=true]{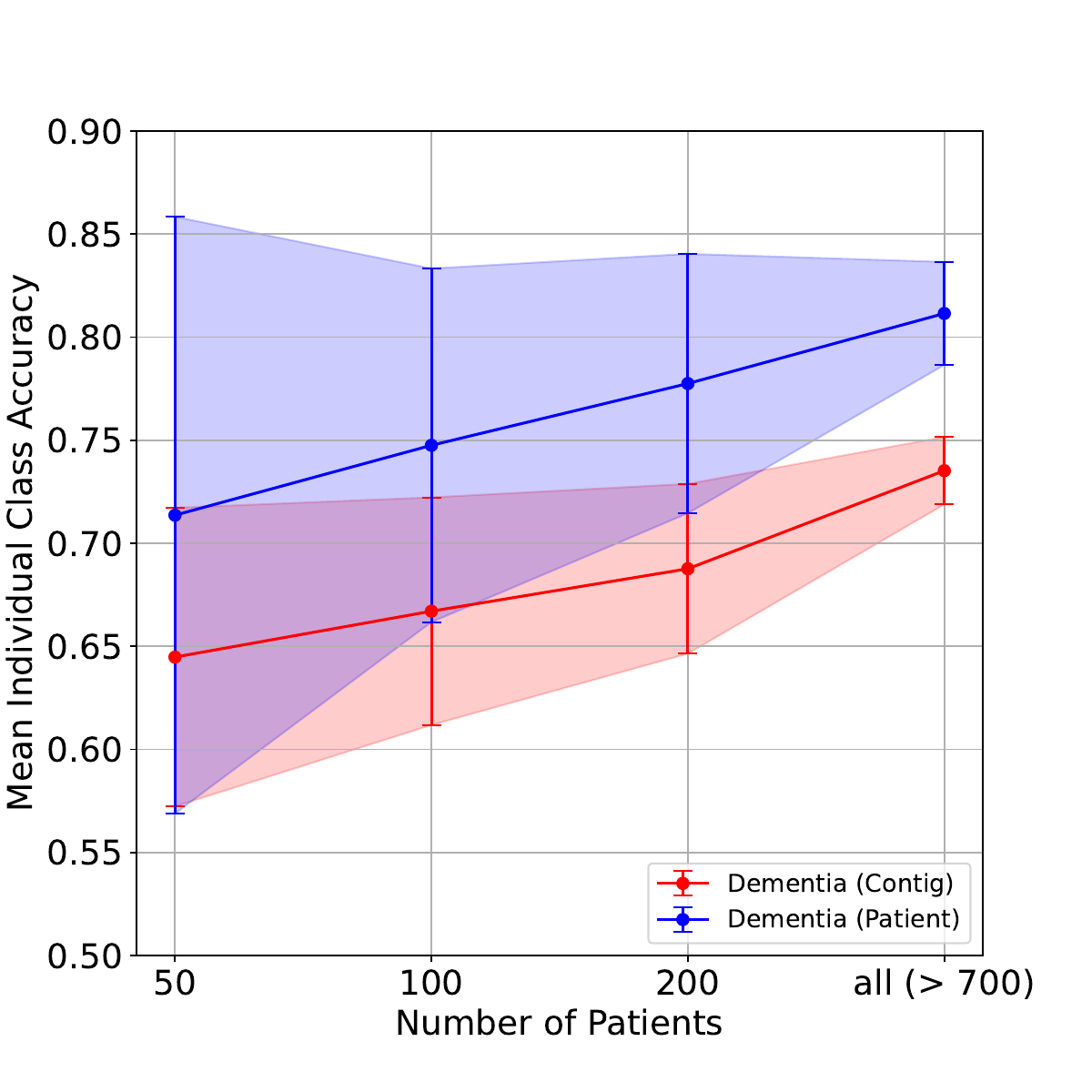}
    \includegraphics[width=0.245\linewidth,trim=0.15cm 0cm 1cm 1cm,clip=true]{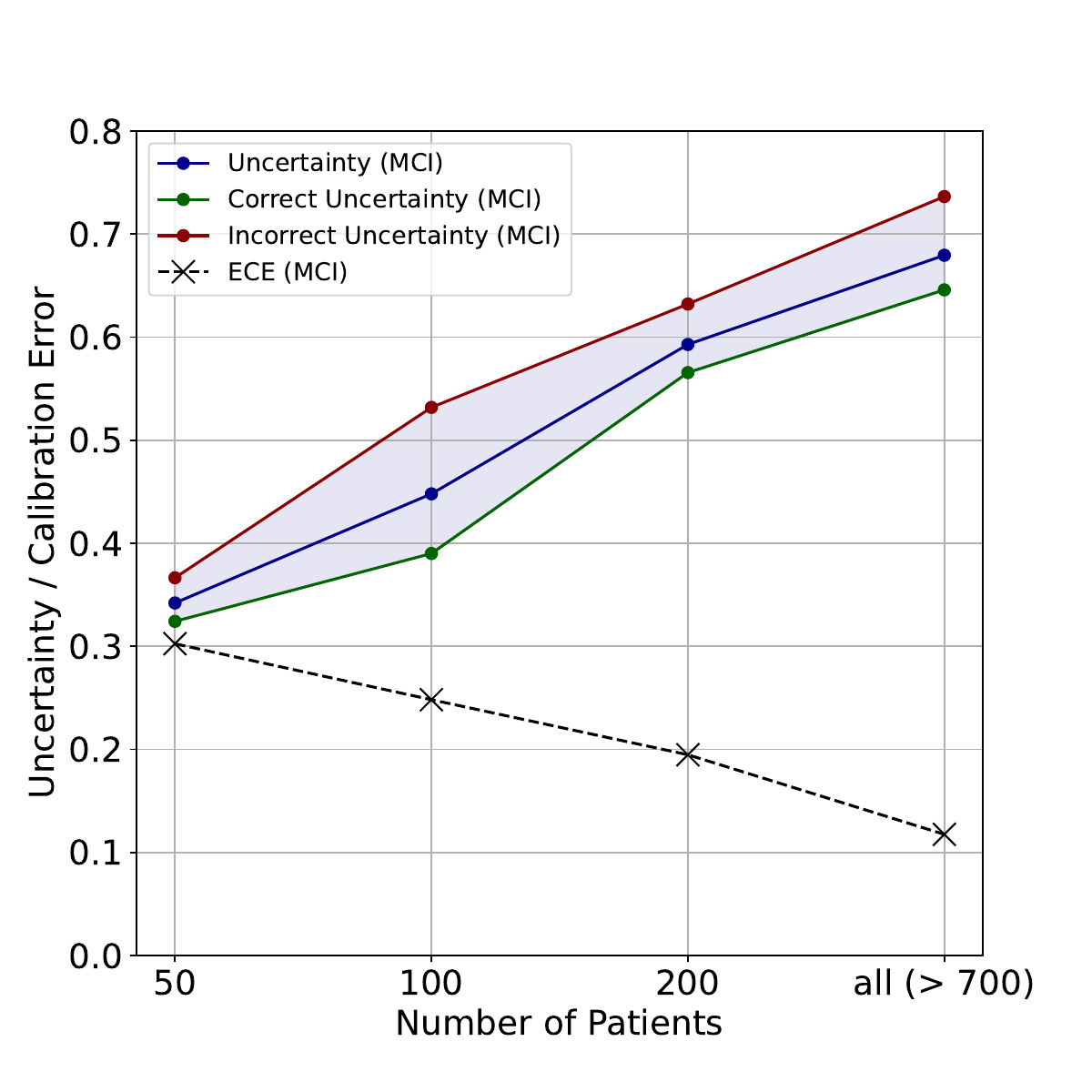}
    \includegraphics[width=0.245\linewidth,trim=0.15cm 0cm 1cm 1cm,clip=true]{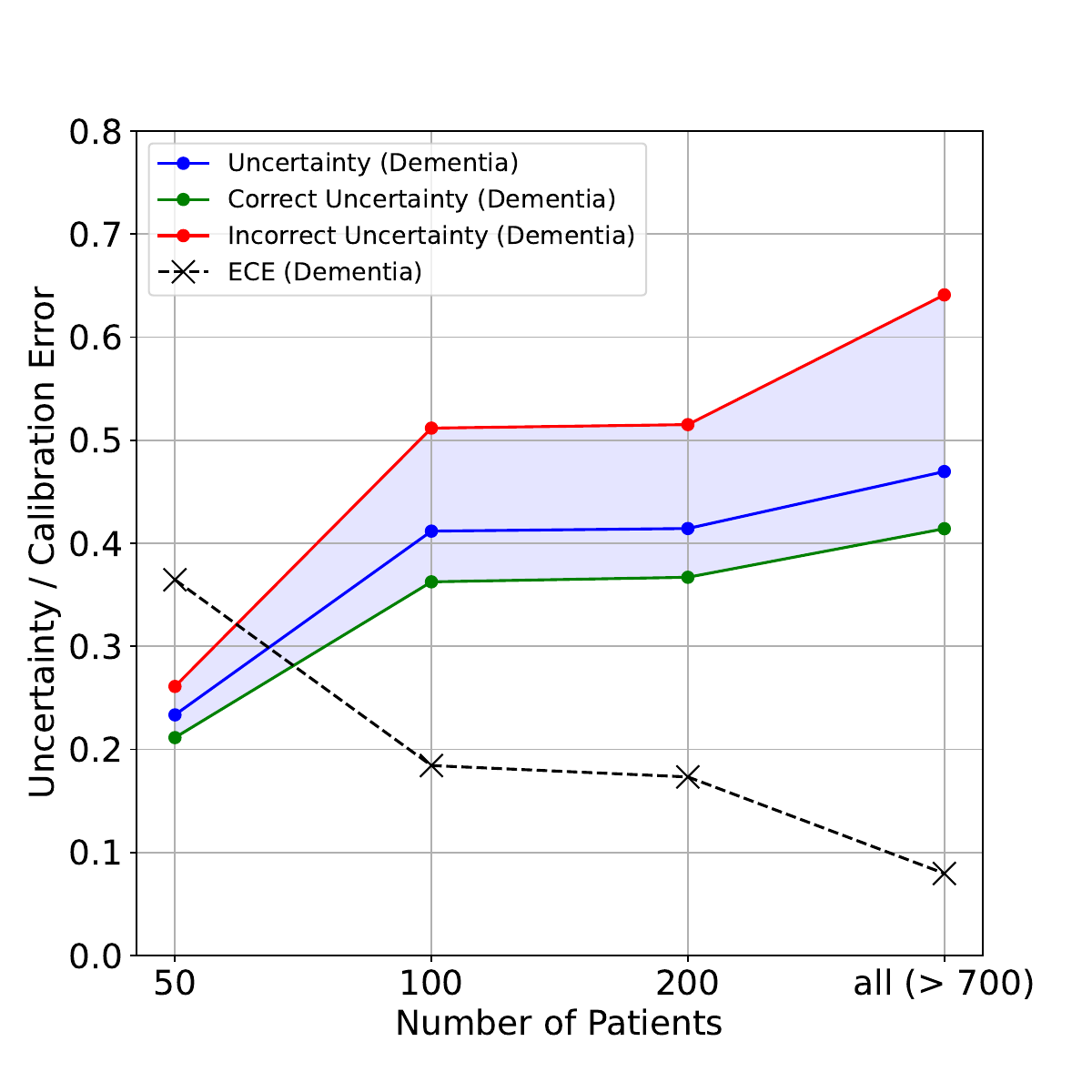}
	\caption{\small{Models get better with less variance in performance as the dataset size increases (first two plots). Moreover, models also become more calibrated with better uncertainty estimates (last two plots).}}\vspace{-10pt}
    \label{fig:metric_evol}
\end{figure}

\paragraph{Issues with Small Datasets} 
From \autoref{tab:main_table}, we see that GENEEG results have a higher variance than CAUEEG results. Models trained on small datasets like GENEEG can fail when a particular train-test split does not simulate \textit{i.i.d. conditions}. To see this effect in CAUEEG, we decided to repeat our experiments on random smaller subsets (of size 50, 100 and 200) of the CAUEEG datasets. 
In \autoref{fig:metric_evol}, we show how Accuracy evolves with increasing dataset size. We used the same model sizes for each dataset size. As the dataset size increases, both contig and patient level test predictions become more accurate and less variant. This implies that the performance of models become more stable as the dataset size increases. 
We also report this trend for reliability metrics. As seen in \autoref{fig:metric_evol}, with increasing dataset size, models become more calibrated, and remarkably, also more uncertain. This implies that models trained on smaller datasets tend to be overconfident in their predictions. Even when they make more errors, they are overconfident about those mistakes. Our analysis calls for caution about the trustworthiness and practical applicability of such systems, especially when the data size is small. This also calls for more effort in quality data collection before making claims of proficient ML-based disease diagnoses using EEG.

\begin{figure*}[t]
    \centering
    \includegraphics[width=0.49\linewidth,trim=2cm 1.5cm 2cm 2cm,clip=true]{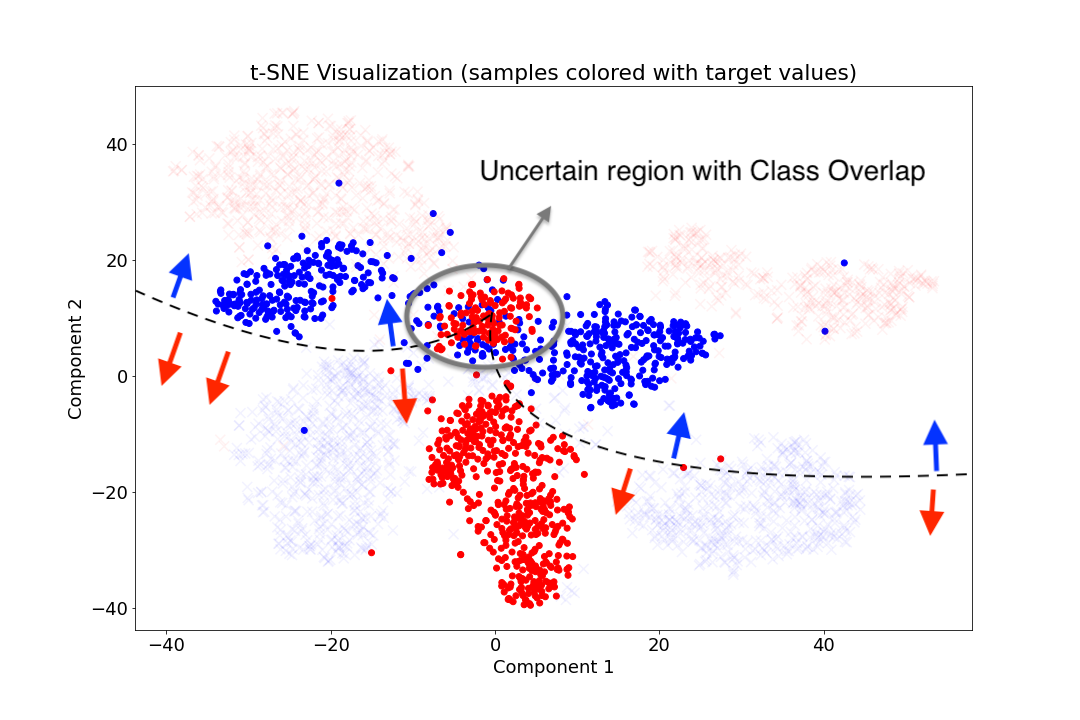}
    \includegraphics[width=0.49\linewidth,trim=2cm 1.5cm 2cm 2cm,clip=true]{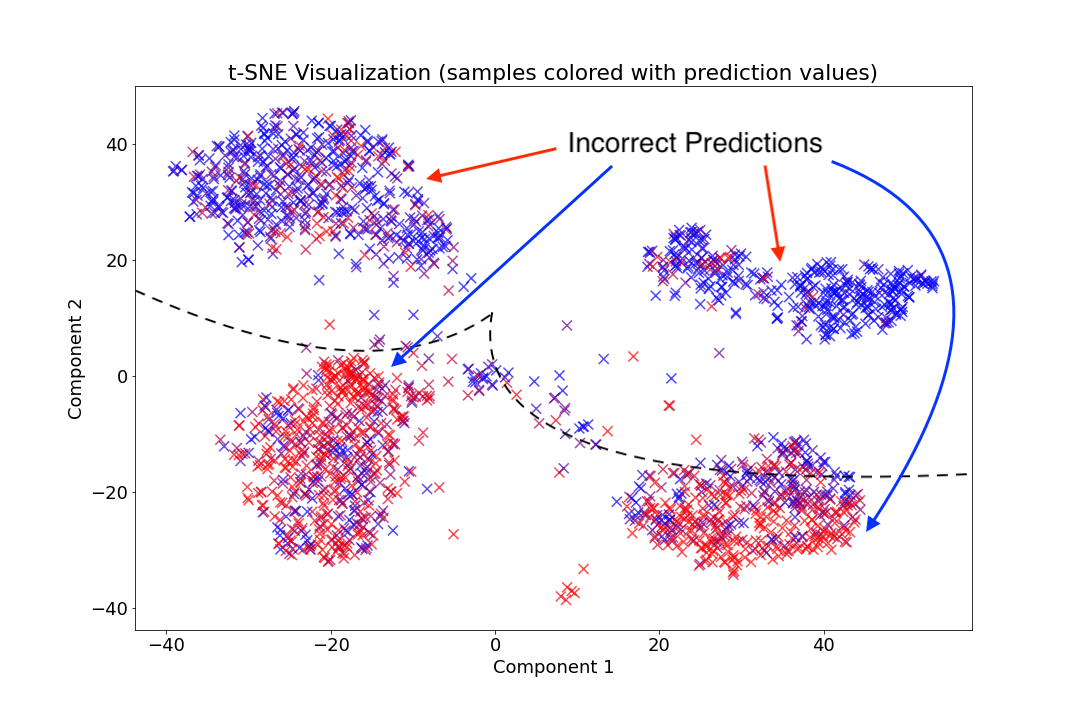}\\
    \small{(a)} \hspace{200pt} \small{(b)}
    \vspace{-5pt}
	\caption{\small{
        (\filledcircr) and (\filledcircb) points in (a) are contigs of some training patients from MCI and Control class respectively. (\ding{53}) represents contigs of some test patients colored similarly (washed out). We see that some patients present contigs that overlap with contigs from the other class, giving rise to \textit{aleatoric uncertainty}. In contrast, some patient contigs may lie in regions lacking training data support reflecting \textit{epistemic uncertainty}. This calls for uncertainty-aware predictors and the optimistically extrapolated decision boundaries (one potential boundary shown with dashed curves) can result in incorrect predictions. 
 }}\vspace{-10pt}
    \label{fig:tsne}
\end{figure*}

\paragraph{Overlap between Classes} 
From the reliability metrics shown in \autoref{fig:metric_evol}, it is clear that as dataset size increases, models become more calibrated. However, \textit{higher calibration in these datasets comes with more uncertainty about unseen test samples.} Moreover, the performance numbers in \autoref{tab:main_table} are not encouraging. Differentiating patients with conditions (especially MCI) from control group patients seems to be hard. To further investigate these phenomena, we looked into data samples from each class. In \autoref{fig:tsne}, we used t-SNE \citep{van2008visualizing} to visualize frequency domain data ($17 \times 26$ dimensional) in two-dimensions. We find that there is overlap between the contig samples from both classes. This points towards aleatoric uncertainty in the data which may be irreducible and explains the low performance numbers in \autoref{tab:main_table}. 

\begin{figure}[b]
    \centering
\includegraphics[width=0.34\linewidth,trim=0cm 0cm 1cm 0.2cm,clip=true]{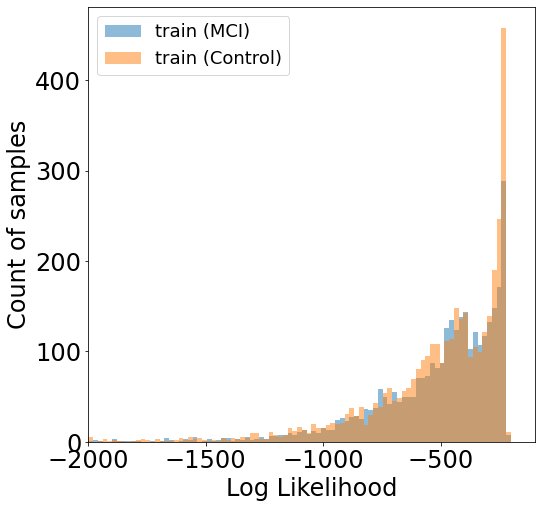}
\includegraphics[width=0.32\linewidth,trim=1cm 0cm 1cm 0.2cm,clip=true]{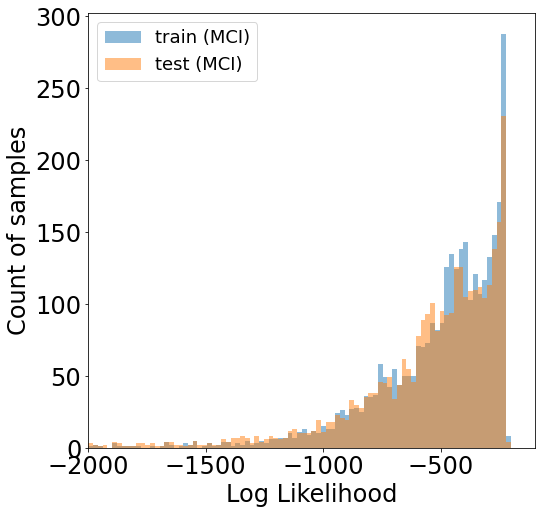}
\includegraphics[width=0.32\linewidth,trim=1cm 0cm 1cm 0.2cm,clip=true]{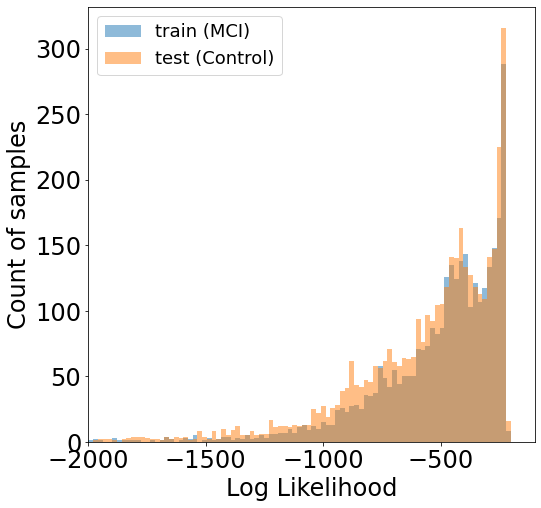}
	\caption{\small{A GMM fit on CAUEEG MCI training set, produces a similar log likelihood distribution on Control training set, MCI test set and Control test set, indicating high overlap.}}\vspace{-20pt}
    \label{fig:density_overlap}
\end{figure}

To show that this overlap exists, not only in these hand-picked t-SNE visualized samples, but in the whole dataset, we use probability density estimation of different subsets of the input dataset. If inputs from different sets are easily separable, the likelihood of samples from different sets should produce a distinct distribution. We fit a Gaussian mixture model (GMM with 10 modes) on the CAUEEG MCI training samples (frequency domain). \autoref{fig:density_overlap} shows the likelihood distributions for data samples in Control training set, MCI test set and Control test set under this trained GMM. We see that the distribution is quite similar among these datasets, which suggests that samples with similar likelihood under the trained density function are distributed similarly in all sets. This highlights the importance of first setting a realistic goal for MCI detection rather than chasing high performance models which may be unreliable. 

\newpage
\begin{wraptable}{r}{0.5\textwidth}
\label{tab:cross_table}
\resizebox{0.5\textwidth}{!}{%
\begin{tabular}{|c|c|}
\hline
\begin{tabular}[c]{@{}c@{}}Training: GENEEG (MCI v Control)\\ Evaluation: CAUEEG (MCI v Control)\end{tabular}      & \begin{tabular}[c]{@{}c@{}}0.41 v 0.65 (Contig)\\ 0.33 v 0.74 (Patient)\end{tabular} \\ \hline
\begin{tabular}[c]{@{}c@{}}Training: GENEEG (MCI v Control)\\ Evaluation: CAUEEG (Dementia v Control)\end{tabular} & \begin{tabular}[c]{@{}c@{}}0.52 v 0.65 (Contig)\\ 0.53 v 0.74 (Patient)\end{tabular} \\ \hline
\begin{tabular}[c]{@{}c@{}}Training: CAUEEG (MCI v Control)\\ Evaluation: GENEEG (MCI v Control)\end{tabular}      & \begin{tabular}[c]{@{}c@{}}0.00 v 1.00 (Contig)\\ 0.00 v 1.00 (Patient)\end{tabular} \\ \hline
\begin{tabular}[c]{@{}c@{}}Training: CAUEEG (Dementia v Control)\\ Evaluation: GENEEG (MCI v Control)\end{tabular} & \begin{tabular}[c]{@{}c@{}}0.00 v 1.00 (Contig)\\ 0.00 v 1.00 (Patient)\end{tabular} \\ \hline
\end{tabular}}
\vspace{-10pt}
\end{wraptable}

\paragraph{Cross-Domain Differences} It is evident from \autoref{tab:main_table} that detecting Dementia is easier compared to detecting MCI in CAUEEG. However, GENEEG MCI detection performance is significantly higher than CAUEEG. 
We suspect that this is because of \textit{differences in data distributions under these two settings}. While clinical evaluations can detect subtle signs of MCI, only more severe symptoms (tending towards Dementia) get diagnosed in general practice clinics making its detection easier.

To test whether these models can handle data \textit{distrbution shifts} across domains, we evaluated CAUEEG models on GENEEG dataset and vice versa. We use the best performing CNN models for this cross-domain testing but see similar results in MLP models. The table above shows all combinations of this testing. We find that:
\begin{itemize}
    \item Models trained on GENEEG (MCI v Control) has significantly worse performance on CAUEEG (MCI v Control) set. While the GENEEG model has a macro patient accuracy of 0.77 on the GENEEG test set, the same model has a macro patient accuracy of 0.54 on the CAUEEG test set. The distribution shift causes a significant drop in performance. 
    \item The same model trained on GENEEG (MCI v Control) performs better on the CAUEEG (Dementia v Control) task, with a macro patient accuracy of 0.64. This supports our hypothesis that \textit{MCI detected in general practice clinics is probably a more severe form tending towards Dementia}. 
    \item CAUEEG models completely fail to detect any GENEEG MCI patients. We suspect that this is caused by a potential \textit{spurious correlation} in the CAUEEG dataset, that does not occur in the GENEEG dataset. We defer the study of this phenomena to future work.
\end{itemize}

\begin{figure*}[t]
    \centering
    \includegraphics[width=0.24\linewidth,trim=1cm 0cm 1cm 1cm,clip=true]{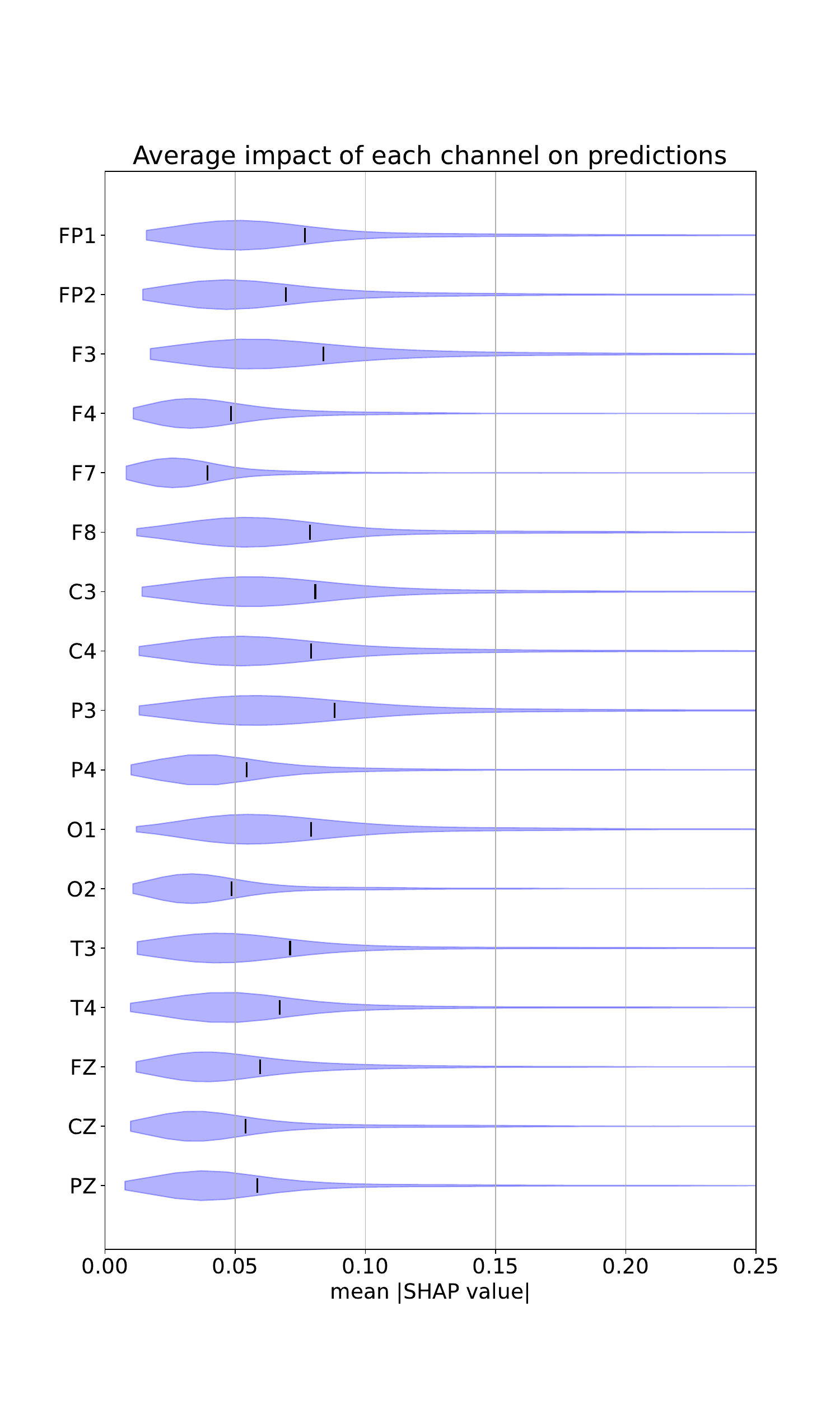}
    \includegraphics[width=0.24\linewidth,trim=1cm 0cm 1cm 1cm,clip=true]{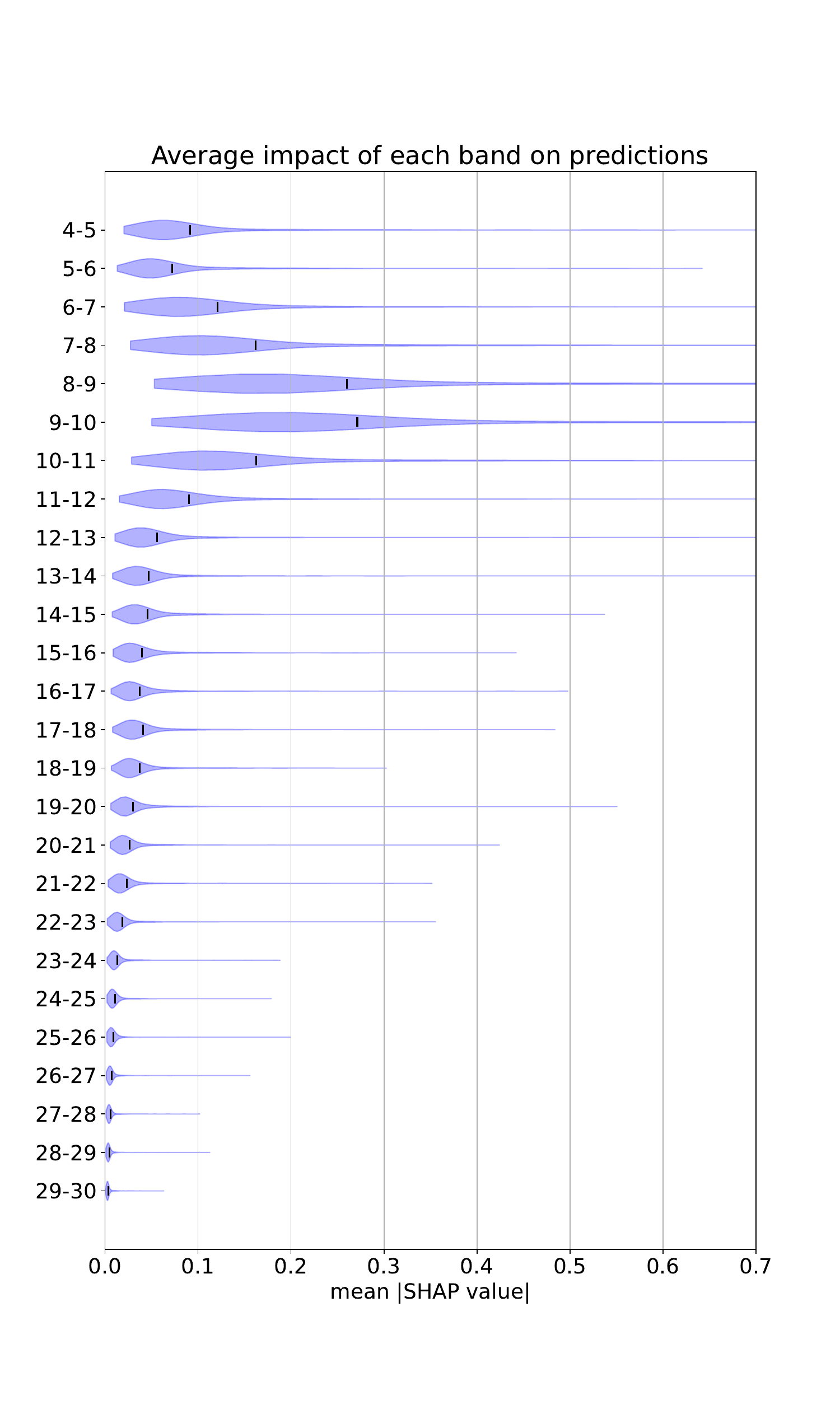}
    \includegraphics[width=0.24\linewidth,trim=1cm 0cm 1cm 1cm,clip=true]{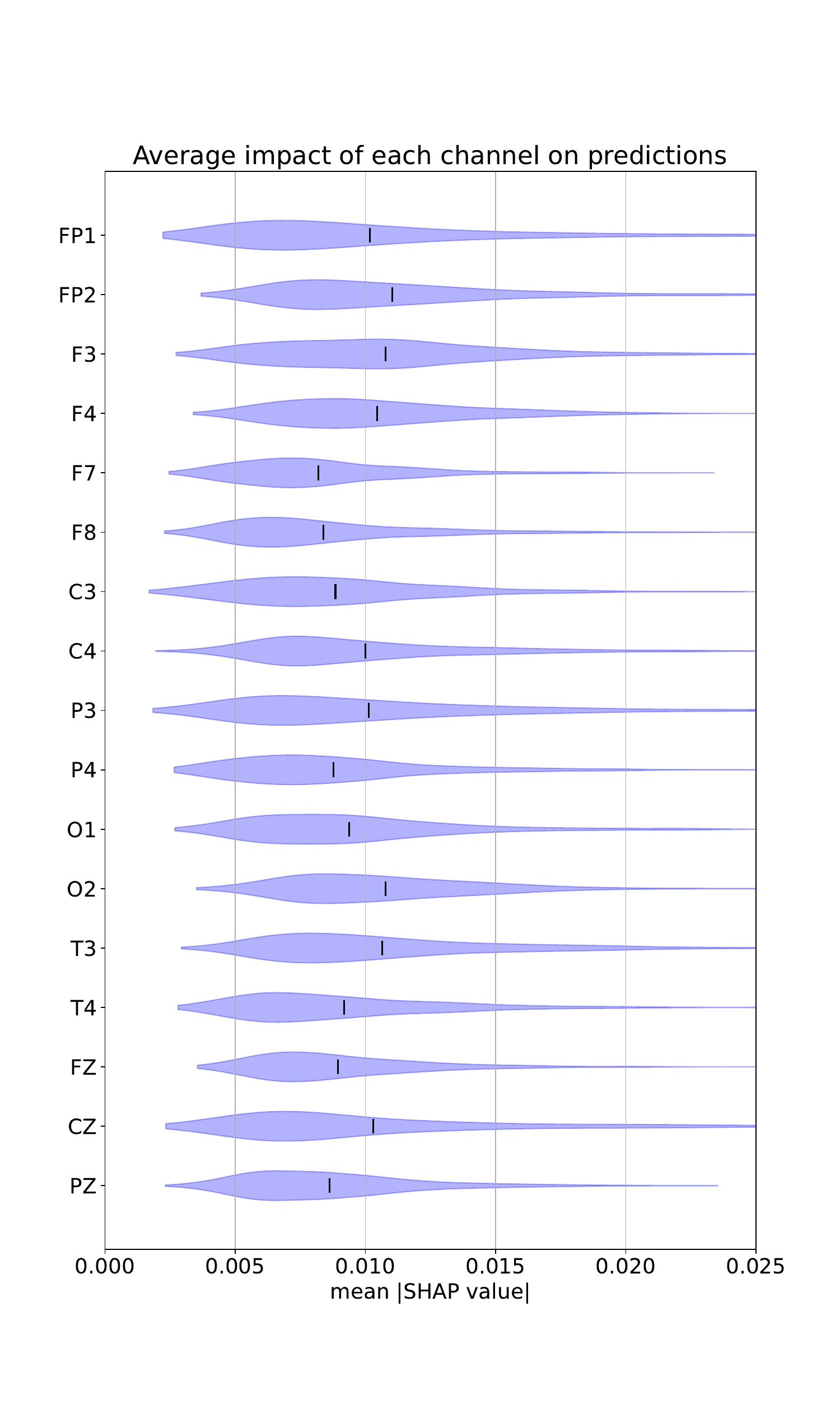}
    \includegraphics[width=0.24\linewidth,trim=1cm 0cm 1cm 1cm,clip=true]{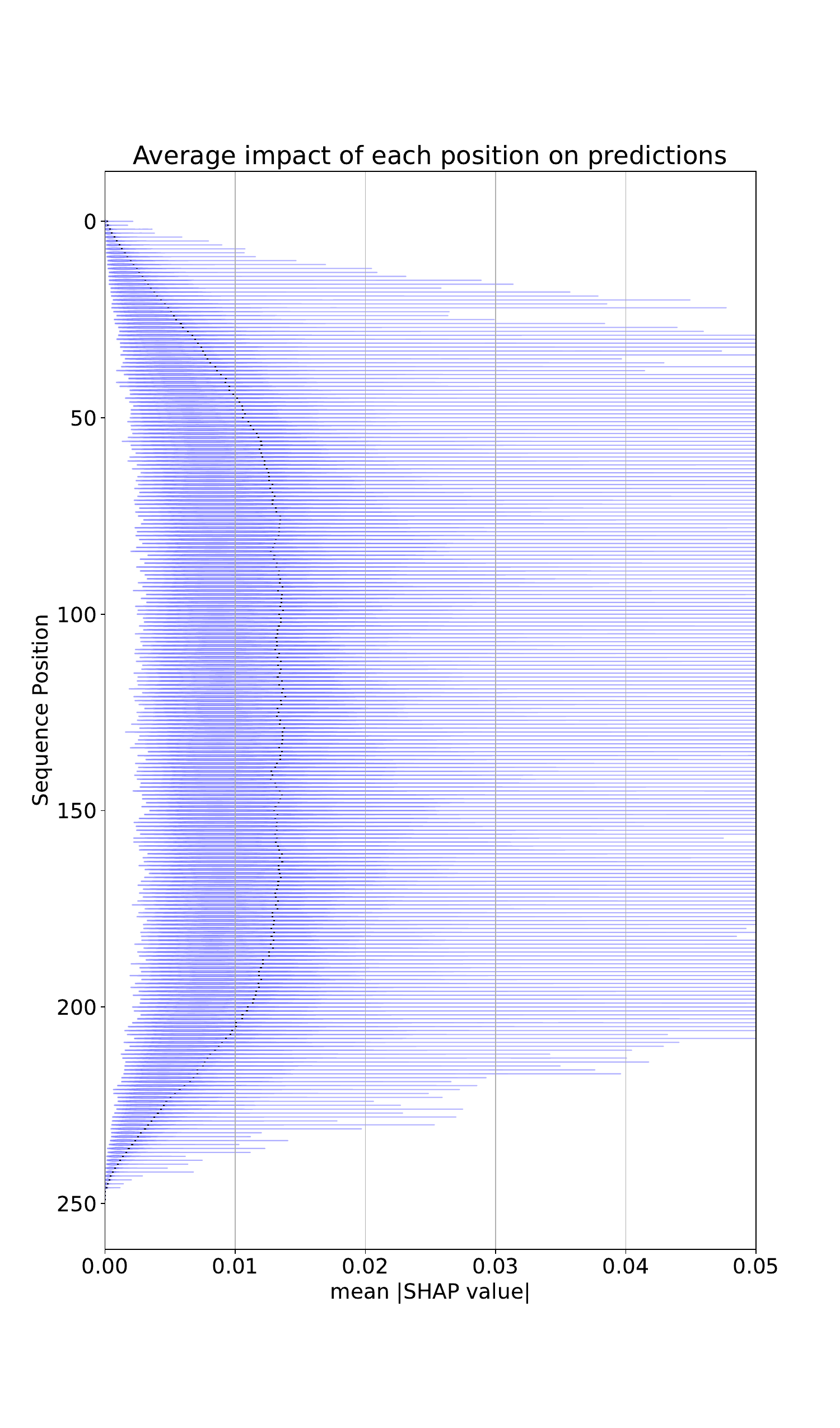}\\
    (a) \hspace{100pt} (b)\hspace{100pt} (c)\hspace{100pt} (d)
	\vspace{-5pt}
	\caption{\small{SHAP value distributions showing the average impact of each feature on model predictions. (a) Frequency-domain channels, (b) Frequency-domain bands, (c) Time-domain channels, (d) Time-domain positions. The SHAP values for (d) are expected and explained in \autoref{par:shap}.}}\vspace{-10pt}
    \label{fig:shap}
\end{figure*}

\section{Additional Discussion}\label{sec:discuss}

\paragraph{Discrepancy in Contig vs Patient level performance}
We see that typically, patient-level performance is higher than the corresponding contig-level performance. This is an artifact of an unequal number of contigs per patient, the contig voting strategy as well as model training. As the majority voting strategy will predict the label by counting all contigs with probabilities between $50 - 100\%$, more mistakes for a patient with more contigs can be compensated by fewer mistakes for a patient with fewer contigs, resulting in the discrepancy. 

\paragraph{Can we remove epistemic uncertainty?} 
From \autoref{fig:metric_evol}, we also see that models trained on larger datasets also tend to become more uncertain about their predictions, while becoming more calibrated. They get a better sense of the epistemic uncertainty present in the data, i.e. regions with less training support are classified with high uncertainty. This is important for reliability, bolstering the need for more efforts in quality data collection. 
Apart from using more high quality data, we can further alleviate \textit{epistemic uncertainty} by making classifiers uncertainty aware. We tried some of the standard methods for uncertainty aware classficiation \citep{liu2020simple, buidensity}, but achieved poor borderline results. In both methods, the density estimation step fails due to high dimensionality.
This results in classifiers being overly conservative and producing almost uniform distribution of probabilities in both classes. We defer further improvement of this approach to future work.

\paragraph{Which features most influence the model's predictions?}\label{par:shap} SHAP values \citep{lundberg2017unified} are a classic way to interpret feature importance for model predictions. They measure the change in model prediction conditioned on a feature. We calculate SHAP values for both domains of data (using MLP and CNN models).
In \autoref{fig:shap}, we show the average impact of each channel, band, and position in making predictions across the test sets of 10 randomly chosen training runs. We can see that for the frequency domain data, low-frequency bands between 6-12 Hz are the most impactful features. Similarly, there is a reasonable spread of importance among channels (P3 being the most impactful, F7 being the least). 
In contrast, time-domain models have a relatively even spread of impact across the channels. Across the temporal positions, the impact distribution exhibited is to be expected. This is because clean contigs generated after artifacting can have patterns spread at any location. Moreover, the impact diminishes towards the ends because of pooling layers in the CNN.

\section{Conclusion and Future Work}
\label{sec:conclusion}

In this work, we find that models for MCI/Dementia detection trained on small datasets 1) are untrustworthy as they make overconfident mistakes and 2) do not generalize between domains. We call for more effort in high quality large scale data collection to tackle these problems. Moreover, we find evidence of aleatoric uncertainty in MCI detection, highlighting the need for setting realistic goals in ML-based MCI diagnosis using EEG. 
In the future, we aim to develop better 
uncertainty-aware learning methods can work better in identifying out-of-distribution samples while retaining performance on samples within the training domain.

\bibliography{jmlr}

\newpage
\appendix

\section{Grid Search}\label{app:grid}

We perform a grid search on different architecture configurations and hyperparameters. In the case of standard MLPs, these configuration choices include the number of layers (depth), size of each layer (width), activation functions, regularization, normalization, etc. For CNNs, there are even choices like size and stride of filters, number of filters in each layer, presence of pooling layers, number of feed-forward layers, etc. For transformers, we experiment with the number of encoder layers, number of heads, and the feed-forward layer dimensions. 

Moreover, once a model configuration is chosen, other training hyper-parameters like the optimizer, learning rates, batch sizes, training iterations, etc. still need to be chosen. Finding the best configuration and hyper-parameters by searching over all possible values is infeasible but has real consequences on the final results. For practical (compute/time constraints) reasons, we restrict our search over some model configurations, learning rates, batch sizes, and training iterations, while keeping the activation functions fixed to ReLU, the optimizer fixed to Adam, and the loss function fixed to binary cross-entropy. By iterating this grid search over 5 different train-test splits, we find configurations for both the CNN and MLP models that worked best for the given data.

We settle on the following variations to be evaluated in our search for the optimal model configurations and training hyper-parameters. 

\begin{enumerate}
    \item For CNN with time-domain data, there is an additional variation as the input data itself changes in size because of varying contig lengths. For contig length $=250$, we search over 3 model configuration: \textit{bigk-many-shallow}: 32,50,1,1,1; 64,50,1,1,1; 128,50,1,1,1; 4,8,1,1,0 \$ 300,24; \textit{mediumk-many-shallow}: 16,20,1,1,1; 32,20,1,1,1; 64,50,1,1,0; 4,20,1,1,1 \$ 332,24; \textit{bigk-many-deep}: 16,50,1,1,0; 32,50,1,1,1; 64,50,1,1,0; 16,8,1,1,0 \$ 320,24. Here, the colon-separated (;) numbers before \$ represent the configuration of conv layers. Each comma-separated number for the layer represents the number of filters, filter size and 0/1 switches for ReLU activation, Batch-Normalization and Max-Pooling layers respectively after the conv layer. Numbers after the \$ represent the flattened dimensionality of the conv layer features and the number of neurons in the hidden feedforward layer respectively. Other parameter choices like varying activation functions, stride of filters, etc are fixed and results from this grid search were used to commit to optimal configurations for other contig length models.  

    \item For MLP with spectral data, we search over 5 model configurations: \textit{wide-shallow}: 512,512; \textit{wide-deep}: 256,256,128,128; \textit{narrow-shallow}: 64,64; \textit{narrow-deep}: 32,32,32,32; \textit{wide-to-narrow}: 128,64,32,16. Here the comma-separated numbers adjacent to each variation represent the number of neurons in that layer. 
    
    \item For transformers with time-domain data, we experiment with the number of heads in the encoder layer, number of encoder layers, and the dimension of the feed-forward network. We try \textit{number of heads: 1, 2, 4, 8}, \textit{number of encoder layers: 1, 4, 12}, \textit{feed-forward network dimension: 128, 256}.  
\end{enumerate}

Apart from the above, the search also considered training hyper-parameters like learning rates (choice between $1e-4, 1e-5$) and batch sizes (choice between $16, 32$). The optimizer is fixed to Adam and the number of training epochs (the number of times the model sees the whole training set) is set to $20$. Mean test accuracy is recorded after each epoch to inform the optimal number of epochs. This process is repeated for 5 different train/test splits, i.e. for each train/test split, models were trained and evaluated for all choices of configurations and hyper-parameters. Using these grid searches, we find that the \textit{bigk-many-shallow} configuration for CNNs (around 50K parameters) and \textit{wide-deep} configuration for MLPs (around 230K parameters) performed best overall. For transformers, we use 4 heads, 4 layers and 256 as the feed-forward layer dimension with around 67K parameters. Along with these, the learning rate $1e-4$, batch-size $16$ and $10$ number of epochs are selected as the optimal hyper-parameters. This also makes the CNN model most parameter efficient but still the most performant.

\begin{table*}[t]
\vspace{-20pt}
\caption{The effect of different pre-processing and other experiment design decisions on model Accuracy \texttt{mean (std)} on the GENEEG dataset. Here, Time Domain corresponds to CNN models. Baseline: $= 0.50$. Bold $=$ better.}
\label{tab:cmacc}
\resizebox{\textwidth}{!}{%
\begin{tabular}{|
>{\columncolor[HTML]{EFEFEF}}c 
>{\columncolor[HTML]{EFEFEF}}c 
>{\columncolor[HTML]{EFEFEF}}c |cccccccc|}
\hline
\multicolumn{1}{|c|}{\cellcolor[HTML]{C0C0C0}} &
  \multicolumn{1}{c|}{\cellcolor[HTML]{C0C0C0}} &
  \cellcolor[HTML]{C0C0C0} &
  \multicolumn{8}{c|}{\cellcolor[HTML]{C0C0C0}Contig Length $\rightarrow$} \\ \cline{4-11} 
\multicolumn{1}{|c|}{\multirow{-2}{*}{\cellcolor[HTML]{C0C0C0}\begin{tabular}[c]{@{}c@{}}Normalization\\ $\downarrow$\end{tabular}}} &
  \multicolumn{1}{c|}{\multirow{-2}{*}{\cellcolor[HTML]{C0C0C0}\begin{tabular}[c]{@{}c@{}}Balancing\\ $\downarrow$\end{tabular}}} &
  \multirow{-2}{*}{\cellcolor[HTML]{C0C0C0}\begin{tabular}[c]{@{}c@{}}Level\\ $\downarrow$\end{tabular}} &
  \multicolumn{2}{c|}{\cellcolor[HTML]{EFEFEF}250} &
  \multicolumn{2}{c|}{\cellcolor[HTML]{EFEFEF}500} &
  \multicolumn{2}{c|}{\cellcolor[HTML]{EFEFEF}1000} &
  \multicolumn{2}{c|}{\cellcolor[HTML]{EFEFEF}2000} \\ \hline
\multicolumn{3}{|c|}{\cellcolor[HTML]{C0C0C0}Domain {[}Time (T) / Frequency (F){]} $\rightarrow$} &
  \multicolumn{1}{c|}{\cellcolor[HTML]{EFEFEF}T} &
  \multicolumn{1}{c|}{\cellcolor[HTML]{EFEFEF}F} &
  \multicolumn{1}{c|}{\cellcolor[HTML]{EFEFEF}T} &
  \multicolumn{1}{c|}{\cellcolor[HTML]{EFEFEF}F} &
  \multicolumn{1}{c|}{\cellcolor[HTML]{EFEFEF}T} &
  \multicolumn{1}{c|}{\cellcolor[HTML]{EFEFEF}F} &
  \multicolumn{1}{c|}{\cellcolor[HTML]{EFEFEF}T} &
  \cellcolor[HTML]{EFEFEF}F \\ \hline
\multicolumn{1}{|c|}{\cellcolor[HTML]{EFEFEF}} &
  \multicolumn{1}{c|}{\cellcolor[HTML]{EFEFEF}} &
  Contig &
  \multicolumn{1}{c|}{\textbf{0.70} (0.05)} &
  \multicolumn{1}{c|}{0.60 (0.04)} &
  \multicolumn{1}{c|}{\textbf{0.73} (0.05)} &
  \multicolumn{1}{c|}{0.62 (0.05)} &
  \multicolumn{1}{c|}{\textbf{0.73} (0.07)} &
  \multicolumn{1}{c|}{0.62 (0.07)} &
  \multicolumn{1}{c|}{\textbf{0.71} (0.07)} &
  0.66 (0.08) \\ \cline{3-11}
\multicolumn{1}{|c|}{\cellcolor[HTML]{EFEFEF}} &
  \multicolumn{1}{c|}{\multirow{-2}{*}{\cellcolor[HTML]{EFEFEF}Balanced}} &
  Patient &
  \multicolumn{1}{c|}{\textbf{0.76} (0.09)} &
  \multicolumn{1}{c|}{0.67 (0.11)} &
  \multicolumn{1}{c|}{\textbf{0.77} (0.07)} &
  \multicolumn{1}{c|}{0.64 (0.10)} &
  \multicolumn{1}{c|}{\textbf{0.75} (0.09)} &
  \multicolumn{1}{c|}{0.64 (0.09)} &
  \multicolumn{1}{c|}{\textbf{0.73} (0.08)} &
  0.67 (0.10) \\ \cline{2-11}
\multicolumn{1}{|c|}{\cellcolor[HTML]{EFEFEF}} &
  \multicolumn{1}{c|}{\cellcolor[HTML]{EFEFEF}} &
  Contig &
  \multicolumn{1}{c|}{\textbf{0.70} (0.05)} &
  \multicolumn{1}{c|}{0.61 (0.05)} &
  \multicolumn{1}{c|}{\textbf{0.75} (0.05)} &
  \multicolumn{1}{c|}{0.65 (0.05)} &
  \multicolumn{1}{c|}{\textbf{0.71} (0.08)} &
  \multicolumn{1}{c|}{0.60 (0.09)} &
  \multicolumn{1}{c|}{\textbf{0.72} (0.09)} &
  0.64 (0.07) \\ \cline{3-11}
\multicolumn{1}{|c|}{\multirow{-4}{*}{\cellcolor[HTML]{EFEFEF}none}} &
  \multicolumn{1}{c|}{\multirow{-2}{*}{\cellcolor[HTML]{EFEFEF}Unbalanced}} &
  Patient &
  \multicolumn{1}{c|}{\textbf{0.75} (0.09)} &
  \multicolumn{1}{c|}{0.68 (0.10)} &
  \multicolumn{1}{c|}{\textbf{0.80} (0.08)} &
  \multicolumn{1}{c|}{0.68 (0.09)} &
  \multicolumn{1}{c|}{\textbf{0.75} (0.10)} &
  \multicolumn{1}{c|}{0.62 (0.11)} &
  \multicolumn{1}{c|}{\textbf{0.74} (0.11)} &
  0.65 (0.08) \\ \hline
\multicolumn{1}{|c|}{\cellcolor[HTML]{EFEFEF}} &
  \multicolumn{1}{c|}{\cellcolor[HTML]{EFEFEF}} &
  Contig &
  \multicolumn{1}{c|}{0.56 (0.08)} &
  \multicolumn{1}{c|}{\textbf{0.61} (0.05)} &
  \multicolumn{1}{c|}{0.61 (0.08)} &
  \multicolumn{1}{c|}{\textbf{0.62} (0.06)} &
  \multicolumn{1}{c|}{0.62 (0.09)} &
  \multicolumn{1}{c|}{\textbf{0.64} (0.08)} &
  \multicolumn{1}{c|}{0.64 (0.11)} &
  \textbf{0.65} (0.07) \\ \cline{3-11}
\multicolumn{1}{|c|}{\cellcolor[HTML]{EFEFEF}} &
  \multicolumn{1}{c|}{\multirow{-2}{*}{\cellcolor[HTML]{EFEFEF}Balanced}} &
  Patient &
  \multicolumn{1}{c|}{0.58 (0.12)} &
  \multicolumn{1}{c|}{\textbf{0.68} (0.12)} &
  \multicolumn{1}{c|}{0.61 (0.11)} &
  \multicolumn{1}{c|}{\textbf{0.66} (0.08)} &
  \multicolumn{1}{c|}{0.63 (0.12)} &
  \multicolumn{1}{c|}{\textbf{0.64} (0.09)} &
  \multicolumn{1}{c|}{0.63 (0.12)} &
  \textbf{0.65} (0.09) \\ \cline{2-11}
\multicolumn{1}{|c|}{\cellcolor[HTML]{EFEFEF}} &
  \multicolumn{1}{c|}{\cellcolor[HTML]{EFEFEF}} &
  Contig &
  \multicolumn{1}{c|}{0.59 (0.08)} &
  \multicolumn{1}{c|}{\textbf{0.62} (0.05)} &
  \multicolumn{1}{c|}{0.62 (0.08)} &
  \multicolumn{1}{c|}{\textbf{0.63} (0.07)} &
  \multicolumn{1}{c|}{0.60 (0.09)} &
  \multicolumn{1}{c|}{\textbf{0.65} (0.09)} &
  \multicolumn{1}{c|}{0.62 (0.10)} &
  \textbf{0.64} (0.09) \\ \cline{3-11}
\multicolumn{1}{|c|}{\multirow{-4}{*}{\cellcolor[HTML]{EFEFEF}minmax}} &
  \multicolumn{1}{c|}{\multirow{-2}{*}{\cellcolor[HTML]{EFEFEF}Unbalanced}} &
  Patient &
  \multicolumn{1}{c|}{0.61 (0.13)} &
  \multicolumn{1}{c|}{\textbf{0.68} (0.11)} &
  \multicolumn{1}{c|}{0.61 (0.10)} &
  \multicolumn{1}{c|}{\textbf{0.67} (0.11)} &
  \multicolumn{1}{c|}{0.61 (0.12)} &
  \multicolumn{1}{c|}{\textbf{0.69} (0.11)} &
  \multicolumn{1}{c|}{0.63 (0.12)} &
  \textbf{0.65} (0.11) \\ \hline
\multicolumn{1}{|c|}{\cellcolor[HTML]{EFEFEF}} &
  \multicolumn{1}{c|}{\cellcolor[HTML]{EFEFEF}} &
  Contig &
  \multicolumn{1}{c|}{\textbf{0.73} (0.05)} &
  \multicolumn{1}{c|}{0.60 (0.05)} &
  \multicolumn{1}{c|}{\textbf{0.74} (0.06)} &
  \multicolumn{1}{c|}{0.62 (0.06)} &
  \multicolumn{1}{c|}{\textbf{0.72} (0.07)} &
  \multicolumn{1}{c|}{0.60 (0.09)} &
  \multicolumn{1}{c|}{\textbf{0.71} (0.08)} &
  0.59 (0.08) \\ \cline{3-11}
\multicolumn{1}{|c|}{\cellcolor[HTML]{EFEFEF}} &
  \multicolumn{1}{c|}{\multirow{-2}{*}{\cellcolor[HTML]{EFEFEF}Balanced}} &
  Patient &
  \multicolumn{1}{c|}{\textbf{0.81} (0.09)} &
  \multicolumn{1}{c|}{0.66 (0.09)} &
  \multicolumn{1}{c|}{\textbf{0.79} (0.09)} &
  \multicolumn{1}{c|}{0.68 (0.09)} &
  \multicolumn{1}{c|}{\textbf{0.75} (0.08)} &
  \multicolumn{1}{c|}{0.60 (0.12)} &
  \multicolumn{1}{c|}{\textbf{0.73} (0.09)} &
  0.59 (0.08) \\ \cline{2-11}
\multicolumn{1}{|c|}{\cellcolor[HTML]{EFEFEF}} &
  \multicolumn{1}{c|}{\cellcolor[HTML]{EFEFEF}} &
  Contig &
  \multicolumn{1}{c|}{\textbf{0.72} (0.04)} &
  \multicolumn{1}{c|}{0.61 (0.04)} &
  \multicolumn{1}{c|}{\textbf{0.72} (0.09)} &
  \multicolumn{1}{c|}{0.61 (0.06)} &
  \multicolumn{1}{c|}{\textbf{0.72} (0.07)} &
  \multicolumn{1}{c|}{0.61 (0.08)} &
  \multicolumn{1}{c|}{\textbf{0.69} (0.07)} &
  0.57 (0.07) \\ \cline{3-11}
\multicolumn{1}{|c|}{\multirow{-4}{*}{\cellcolor[HTML]{EFEFEF}meanstd}} &
  \multicolumn{1}{c|}{\multirow{-2}{*}{\cellcolor[HTML]{EFEFEF}Unbalanced}} &
  Patient &
  \multicolumn{1}{c|}{\textbf{0.79} (0.07)} &
  \multicolumn{1}{c|}{0.70 (0.09)} &
  \multicolumn{1}{c|}{\textbf{0.76} (0.11)} &
  \multicolumn{1}{c|}{0.63 (0.11)} &
  \multicolumn{1}{c|}{\textbf{0.75} (0.10)} &
  \multicolumn{1}{c|}{0.63 (0.12)} &
  \multicolumn{1}{c|}{\textbf{0.72} (0.08)} &
  0.57 (0.08) \\ \hline

\end{tabular}%
}
\end{table*}

\end{document}